\documentstyle[12pt,epsfig,a4,cite,epic,ecltree,subfigure,fancybox,floatflt,multicol,mon,acronym,qsymbols]{article}
\newcommand{\snu}{\tilde{\nu}}

\newcommand{\gevc}{\mbox{GeV$/c$}}
\newcommand{\gevcc}{\mbox{GeV$/c^2$}}

\begin{document}
\normalsize
\newpage
\pagestyle{plain}
\setcounter{page}{1}
\pagenumbering{arabic}
\normalsize
\setlength{\topmargin}{0cm}
\setlength{\oddsidemargin}{0cm}
 
\font\eightrm=cmr8
\font\ninerm=cmr9
\date{}
\title{ \null\vspace{1cm}
{\Large \bf {Search for $\boldmath R$-Parity Violating Production\\
of Single Sneutrinos\\
in $\bf e^+e^-$ Collisions at $\boldmath \sqrt{s}$ = 189--209 GeV}}
\vspace{1cm}}
\author{The ALEPH Collaboration$^*)$}

\maketitle

\begin{picture}(160,1)
\put(-1,115){\large EUROPEAN ORGANIZATION FOR NUCLEAR RESEARCH (CERN)}
\put(125,94){\parbox[t]{45mm}{\tt CERN-EP/2001-094}}
\put(125,88){\parbox[t]{45mm}{\tt 17-Dec-2001}}
\end{picture}

\vspace{.2cm}
\begin{abstract}
\vspace{.2cm}
   A search for single sneutrino production under the assumption that
$R$-parity is violated via a single dominant $LL\bar{E}$ coupling is 
presented. This search considers the process
${\rm e} \gamma \;{\smash{\mathop{\rightarrow}}}\;\tilde{\nu}\ell$ 
and is performed using the data collected by the ALEPH detector at 
centre-of-mass energies from 189\,GeV up to 209\,GeV corresponding
to an integrated luminosity of 637.1\,$\mathrm{pb}^{-1}$. 
The numbers of observed candidate events 
are in agreement with Standard Model expectations 
and 95\% confidence level upper limits on five of the $LL\bar{E}$ couplings 
are given as a function of the assumed sneutrino mass.
\end{abstract}

\vfill
\centerline{\it Submitted to European Physical Journal C}
\vskip .5cm
\noindent
--------------------------------------------\hfil\break
{\ninerm $^*)$ See next pages for the list of authors}

\newpage
\pagestyle{empty}
\newpage
\small
%
%
\newlength{\saveparskip}
\newlength{\savetextheight}
\newlength{\savetopmargin}
\newlength{\savetextwidth}
\newlength{\saveoddsidemargin}
\newlength{\savetopsep}
\setlength{\saveparskip}{\parskip}
\setlength{\savetextheight}{\textheight}
\setlength{\savetopmargin}{\topmargin}
\setlength{\savetextwidth}{\textwidth}
\setlength{\saveoddsidemargin}{\oddsidemargin}
\setlength{\savetopsep}{\topsep}
%
%
\setlength{\parskip}{0.0cm}
\setlength{\textheight}{25.0cm}
\setlength{\topmargin}{-1.5cm}
\setlength{\textwidth}{16 cm}
\setlength{\oddsidemargin}{-0.0cm}
\setlength{\topsep}{1mm}
\pretolerance=10000
\centerline{\large\bf The ALEPH Collaboration}
\footnotesize
\vspace{0.5cm}
{\raggedbottom
\begin{sloppypar}
\samepage\noindent
A.~Heister,
S.~Schael
\nopagebreak
\begin{center}
\parbox{15.5cm}{\sl\samepage
Physikalisches Institut das RWTH-Aachen, D-52056 Aachen, Germany}
\end{center}\end{sloppypar}
\vspace{2mm}
\begin{sloppypar}
\noindent
R.~Barate,
R.~Bruneli\`ere,
I.~De~Bonis,
D.~Decamp,
C.~Goy,
S.~Jezequel,
J.-P.~Lees,
F.~Martin,
E.~Merle,
\mbox{M.-N.~Minard},
B.~Pietrzyk,
B.~Trocm\'e
\nopagebreak
\begin{center}
\parbox{15.5cm}{\sl\samepage
Laboratoire de Physique des Particules (LAPP), IN$^{2}$P$^{3}$-CNRS,
F-74019 Annecy-le-Vieux Cedex, France}
\end{center}\end{sloppypar}
\vspace{2mm}
\begin{sloppypar}
\noindent
G.~Boix,
S.~Bravo,
M.P.~Casado,
M.~Chmeissani,
J.M.~Crespo,
E.~Fernandez,
M.~Fernandez-Bosman,
Ll.~Garrido,$^{15}$
E.~Graug\'{e}s,
J.~Lopez,
M.~Martinez,
G.~Merino,
R.~Miquel,$^{31}$
Ll.M.~Mir,$^{31}$
A.~Pacheco,
D.~Paneque,
H.~Ruiz
\nopagebreak
\begin{center}
\parbox{15.5cm}{\sl\samepage
Institut de F\'{i}sica d'Altes Energies, Universitat Aut\`{o}noma
de Barcelona, E-08193 Bellaterra (Barcelona), Spain$^{7}$}
\end{center}\end{sloppypar}
\vspace{2mm}
\begin{sloppypar}
\noindent
A.~Colaleo,
D.~Creanza,
N.~De~Filippis,
M.~de~Palma,
G.~Iaselli,
G.~Maggi,
M.~Maggi,
S.~Nuzzo,
A.~Ranieri,
G.~Raso,$^{24}$
F.~Ruggieri,
G.~Selvaggi,
L.~Silvestris,
P.~Tempesta,
A.~Tricomi,$^{3}$
G.~Zito
\nopagebreak
\begin{center}
\parbox{15.5cm}{\sl\samepage
Dipartimento di Fisica, INFN Sezione di Bari, I-70126 Bari, Italy}
\end{center}\end{sloppypar}
\vspace{2mm}
\begin{sloppypar}
\noindent
X.~Huang,
J.~Lin,
Q. Ouyang,
T.~Wang,
Y.~Xie,
R.~Xu,
S.~Xue,
J.~Zhang,
L.~Zhang,
W.~Zhao
\nopagebreak
\begin{center}
\parbox{15.5cm}{\sl\samepage
Institute of High Energy Physics, Academia Sinica, Beijing, The People's
Republic of China$^{8}$}
\end{center}\end{sloppypar}
\vspace{2mm}
\begin{sloppypar}
\noindent
D.~Abbaneo,
P.~Azzurri,
T.~Barklow,$^{30}$
O.~Buchm\"uller,$^{30}$
M.~Cattaneo,
F.~Cerutti,
B.~Clerbaux,
H.~Drevermann,
R.W.~Forty,
M.~Frank,
F.~Gianotti,
T.C.~Greening,$^{26}$
J.B.~Hansen,
J.~Harvey,
D.E.~Hutchcroft,
P.~Janot,
B.~Jost,
M.~Kado,$^{31}$
P.~Maley,
P.~Mato,
A.~Moutoussi,
F.~Ranjard,
L.~Rolandi,
D.~Schlatter,
G.~Sguazzoni,
W.~Tejessy,
F.~Teubert,
A.~Valassi,
I.~Videau,
J.J.~Ward
\nopagebreak
\begin{center}
\parbox{15.5cm}{\sl\samepage
European Laboratory for Particle Physics (CERN), CH-1211 Geneva 23,
Switzerland}
\end{center}\end{sloppypar}
\vspace{2mm}
\begin{sloppypar}
\noindent
F.~Badaud,
S.~Dessagne,
A.~Falvard,$^{20}$
D.~Fayolle,
P.~Gay,
J.~Jousset,
B.~Michel,
S.~Monteil,
D.~Pallin,
J.M.~Pascolo,
P.~Perret
\nopagebreak
\begin{center}
\parbox{15.5cm}{\sl\samepage
Laboratoire de Physique Corpusculaire, Universit\'e Blaise Pascal,
IN$^{2}$P$^{3}$-CNRS, Clermont-Ferrand, F-63177 Aubi\`{e}re, France}
\end{center}\end{sloppypar}
\vspace{2mm}
\begin{sloppypar}
\noindent
J.D.~Hansen,
J.R.~Hansen,
P.H.~Hansen,
B.S.~Nilsson,
A.~W\"a\"an\"anen
\nopagebreak
\begin{center}
\parbox{15.5cm}{\sl\samepage
Niels Bohr Institute, 2100 Copenhagen, DK-Denmark$^{9}$}
\end{center}\end{sloppypar}
\vspace{2mm}
\begin{sloppypar}
\noindent
A.~Kyriakis,
C.~Markou,
E.~Simopoulou,
A.~Vayaki,
K.~Zachariadou
\nopagebreak
\begin{center}
\parbox{15.5cm}{\sl\samepage
Nuclear Research Center Demokritos (NRCD), GR-15310 Attiki, Greece}
\end{center}\end{sloppypar}
\vspace{2mm}
\begin{sloppypar}
\noindent
A.~Blondel,$^{12}$
\mbox{J.-C.~Brient},
F.~Machefert,
A.~Roug\'{e},
M.~Swynghedauw,
R.~Tanaka
\linebreak
H.~Videau
\nopagebreak
\begin{center}
\parbox{15.5cm}{\sl\samepage
Laboratoire de Physique Nucl\'eaire et des Hautes Energies, Ecole
Polytechnique, IN$^{2}$P$^{3}$-CNRS, \mbox{F-91128} Palaiseau Cedex, France}
\end{center}\end{sloppypar}
\vspace{2mm}
\begin{sloppypar}
\noindent
V.~Ciulli,
E.~Focardi,
G.~Parrini
\nopagebreak
\begin{center}
\parbox{15.5cm}{\sl\samepage
Dipartimento di Fisica, Universit\`a di Firenze, INFN Sezione di Firenze,
I-50125 Firenze, Italy}
\end{center}\end{sloppypar}
\vspace{2mm}
\begin{sloppypar}
\noindent
A.~Antonelli,
M.~Antonelli,
G.~Bencivenni,
G.~Bologna,$^{4}$
F.~Bossi,
P.~Campana,
G.~Capon,
V.~Chiarella,
P.~Laurelli,
G.~Mannocchi,$^{5}$
F.~Murtas,
G.P.~Murtas,
L.~Passalacqua,
M.~Pepe-Altarelli,$^{25}$
P.~Spagnolo
\nopagebreak
\begin{center}
\parbox{15.5cm}{\sl\samepage
Laboratori Nazionali dell'INFN (LNF-INFN), I-00044 Frascati, Italy}
\end{center}\end{sloppypar}
\vspace{2mm}
\begin{sloppypar}
\noindent
J.~Kennedy,
J.G.~Lynch,
P.~Negus,
V.~O'Shea,
D.~Smith,
A.S.~Thompson
\nopagebreak
\begin{center}
\parbox{15.5cm}{\sl\samepage
Department of Physics and Astronomy, University of Glasgow, Glasgow G12
8QQ,United Kingdom$^{10}$}
\end{center}\end{sloppypar}
\vspace{2mm}
\begin{sloppypar}
\noindent
\newpage
\noindent
S.~Wasserbaech
\nopagebreak
\begin{center}
\parbox{15.5cm}{\sl\samepage
Department of Physics, Haverford College, Haverford, PA 19041-1392, U.S.A.}
\end{center}\end{sloppypar}
\vspace{2mm}
\begin{sloppypar}
\noindent
R.~Cavanaugh,
S.~Dhamotharan,
C.~Geweniger,
P.~Hanke,
V.~Hepp,
E.E.~Kluge,
G.~Leibenguth,
A.~Putzer,
H.~Stenzel,
K.~Tittel,
S.~Werner,$^{19}$
M.~Wunsch$^{19}$
\nopagebreak
\begin{center}
\parbox{15.5cm}{\sl\samepage
Kirchhoff-Institut f\"ur Physik, Universit\"at Heidelberg, D-69120
Heidelberg, Germany$^{16}$}
\end{center}\end{sloppypar}
\vspace{2mm}
\begin{sloppypar}
\noindent
R.~Beuselinck,
D.M.~Binnie,
W.~Cameron,
G.~Davies,
P.J.~Dornan,
M.~Girone,$^{1}$
R.D.~Hill,
N.~Marinelli,
J.~Nowell,
H.~Przysiezniak,$^{2}$
S.A.~Rutherford,
J.K.~Sedgbeer,
J.C.~Thompson,$^{14}$
R.~White
\nopagebreak
\begin{center}
\parbox{15.5cm}{\sl\samepage
Department of Physics, Imperial College, London SW7 2BZ,
United Kingdom$^{10}$}
\end{center}\end{sloppypar}
\vspace{2mm}
\begin{sloppypar}
\noindent
V.M.~Ghete,
P.~Girtler,
E.~Kneringer,
D.~Kuhn,
G.~Rudolph
\nopagebreak
\begin{center}
\parbox{15.5cm}{\sl\samepage
Institut f\"ur Experimentalphysik, Universit\"at Innsbruck, A-6020
Innsbruck, Austria$^{18}$}
\end{center}\end{sloppypar}
\vspace{2mm}
\begin{sloppypar}
\noindent
E.~Bouhova-Thacker,
C.K.~Bowdery,
D.P.~Clarke,
G.~Ellis,
A.J.~Finch,
F.~Foster,
G.~Hughes,
R.W.L.~Jones,
M.R.~Pearson,
N.A.~Robertson,
M.~Smizanska
\nopagebreak
\begin{center}
\parbox{15.5cm}{\sl\samepage
Department of Physics, University of Lancaster, Lancaster LA1 4YB,
United Kingdom$^{10}$}
\end{center}\end{sloppypar}
\vspace{2mm}
\begin{sloppypar}
\noindent
V.~Lemaitre
\nopagebreak
\begin{center}
\parbox{15.5cm}{\sl\samepage
Institut de Physique Nucl\'eaire, D\'epartement de Physique, Universit\'e Catholique de Louvain, 1348 Louvain-la-Neuve, Belgium}
\end{center}\end{sloppypar}
\vspace{2mm}
\begin{sloppypar}
\noindent
U.~Blumenschein,
F.~H\"olldorfer,
K.~Jakobs,
F.~Kayser,
K.~Kleinknecgt,
A.-S.~M\"uller,
G.~Quast,$^{6}$
B.~Renk,
H.-G.~Sander,
S.~Schmeling,
H.~Wachsmuth,
C.~Zeitnitz,
T.~Ziegler
\nopagebreak
\begin{center}
\parbox{15.5cm}{\sl\samepage
Institut f\"ur Physik, Universit\"at Mainz, D-55099 Mainz, Germany$^{16}$}
\end{center}\end{sloppypar}
\vspace{2mm}
\begin{sloppypar}
\noindent
A.~Bonissent,
J.~Carr,
P.~Coyle,
C.~Curtil,
A.~Ealet,
D.~Fouchez,
O.~Leroy,
T.~Kachelhoffer,
P.~Payre,
D.~Rousseau,
A.~Tilquin
\nopagebreak
\begin{center}
\parbox{15.5cm}{\sl\samepage
Centre de Physique des Particules de Marseille, Univ M\'editerran\'ee,
IN$^{2}$P$^{3}$-CNRS, F-13288 Marseille, France}
\end{center}\end{sloppypar}
\vspace{2mm}
\begin{sloppypar}
\noindent
F.~Ragusa
\nopagebreak
\begin{center}
\parbox{15.5cm}{\sl\samepage
Dipartimento di Fisica, Universit\`a di Milano e INFN Sezione di
Milano, I-20133 Milano, Italy.}
\end{center}\end{sloppypar}
\vspace{2mm}
\begin{sloppypar}
\noindent
A.~David,
H.~Dietl,
G.~Ganis,$^{27}$
K.~H\"uttmann,
G.~L\"utjens,
C.~Mannert,
W.~M\"anner,
\mbox{H.-G.~Moser},
R.~Settles,
G.~Wolf
\nopagebreak
\begin{center}
\parbox{15.5cm}{\sl\samepage
Max-Planck-Institut f\"ur Physik, Werner-Heisenberg-Institut,
D-80805 M\"unchen, Germany\footnotemark[16]}
\end{center}\end{sloppypar}
\vspace{2mm}
\begin{sloppypar}
\noindent
J.~Boucrot,
O.~Callot,
M.~Davier,
L.~Duflot,
\mbox{J.-F.~Grivaz},
Ph.~Heusse,
A.~Jacholkowska,$^{20}$
C.~Loomis,
L.~Serin,
\mbox{J.-J.~Veillet},
J.-B.~de~Vivie~de~R\'egie,$^{28}$
C.~Yuan
\nopagebreak
\begin{center}
\parbox{15.5cm}{\sl\samepage
Laboratoire de l'Acc\'el\'erateur Lin\'eaire, Universit\'e de Paris-Sud,
IN$^{2}$P$^{3}$-CNRS, F-91898 Orsay Cedex, France}
\end{center}\end{sloppypar}
\vspace{2mm}
\begin{sloppypar}
\noindent
G.~Bagliesi,
T.~Boccali,
L.~Fo\`a,
A.~Giammanco,
A.~Giassi,
F.~Ligabue,
A.~Messineo,
F.~Palla,
G.~Sanguinetti,
A.~Sciab\`a,
R.~Tenchini,$^{1}$
A.~Venturi,$^{1}$
P.G.~Verdini
\samepage
\begin{center}
\parbox{15.5cm}{\sl\samepage
Dipartimento di Fisica dell'Universit\`a, INFN Sezione di Pisa,
e Scuola Normale Superiore, I-56010 Pisa, Italy}
\end{center}\end{sloppypar}
\vspace{2mm}
\begin{sloppypar}
\noindent
O.~Awunor,
G.A.~Blair,
J.~Coles,
G.~Cowan,
A.~Garcia-Bellido,
M.G.~Green,
L.T.~Jones,
T.~Medcalf,
A.~Misiejuk,
J.A.~Strong,
P.~Teixeira-Dias
\nopagebreak
\begin{center}
\parbox{15.5cm}{\sl\samepage
Department of Physics, Royal Holloway \& Bedford New College,
University of London, Egham, Surrey TW20 OEX, United Kingdom$^{10}$}
\end{center}\end{sloppypar}
\vspace{2mm}
\begin{sloppypar}
\noindent
R.W.~Clifft,
T.R.~Edgecock,
P.R.~Norton,
I.R.~Tomalin
\nopagebreak
\begin{center}
\parbox{15.5cm}{\sl\samepage
Particle Physics Dept., Rutherford Appleton Laboratory,
Chilton, Didcot, Oxon OX11 OQX, United Kingdom$^{10}$}
\end{center}\end{sloppypar}
\vspace{2mm}
\begin{sloppypar}
\noindent
\mbox{B.~Bloch-Devaux},
D.~Boumediene,
P.~Colas,
B.~Fabbro,
E.~Lan\c{c}on,
\mbox{M.-C.~Lemaire},
E.~Locci,
P.~Perez,
J.~Rander,
\mbox{J.-F.~Renardy},
A.~Rosowsky,
P.~Seager,$^{13}$
A.~Trabelsi,$^{21}$
B.~Tuchming,
B.~Vallage
\nopagebreak
\begin{center}
\parbox{15.5cm}{\sl\samepage
CEA, DAPNIA/Service de Physique des Particules,
CE-Saclay, F-91191 Gif-sur-Yvette Cedex, France$^{17}$}
\end{center}\end{sloppypar}
\newpage
\null\vskip -2cm
\begin{sloppypar}
\noindent
N.~Konstantinidis,
A.M.~Litke,
G.~Taylor
\nopagebreak
\begin{center}
\parbox{15.5cm}{\sl\samepage
Institute for Particle Physics, University of California at
Santa Cruz, Santa Cruz, CA 95064, USA$^{22}$}
\end{center}\end{sloppypar}
\vspace{2mm}
\begin{sloppypar}
\noindent
C.N.~Booth,
S.~Cartwright,
F.~Combley,$^{4}$
P.N.~Hodgson,
M.~Lehto,
L.F.~Thompson
\nopagebreak
\begin{center}
\parbox{15.5cm}{\sl\samepage
Department of Physics, University of Sheffield, Sheffield S3 7RH,
United Kingdom$^{10}$}
\end{center}\end{sloppypar}
\vspace{2mm}
\begin{sloppypar}
\noindent
K.~Affholderbach,$^{23}$
A.~B\"ohrer,
S.~Brandt,
C.~Grupen,
J.~Hess,
A.~Ngac,
G.~Prange,
U.~Sieler
\nopagebreak
\begin{center}
\parbox{15.5cm}{\sl\samepage
Fachbereich Physik, Universit\"at Siegen, D-57068 Siegen, Germany$^{16}$}
\end{center}\end{sloppypar}
\vspace{2mm}
\begin{sloppypar}
\noindent
C.~Borean,
G.~Giannini
\nopagebreak
\begin{center}
\parbox{15.5cm}{\sl\samepage
Dipartimento di Fisica, Universit\`a di Trieste e INFN Sezione di Trieste,
I-34127 Trieste, Italy}
\end{center}\end{sloppypar}
\vspace{2mm}
\begin{sloppypar}
\noindent
H.~He,
J.~Putz,
J.~Rothberg
\nopagebreak
\begin{center}
\parbox{15.5cm}{\sl\samepage
Experimental Elementary Particle Physics, University of Washington, Seattle,
WA 98195 U.S.A.}
\end{center}\end{sloppypar}
\vspace{2mm}
\begin{sloppypar}
\noindent
S.R.~Armstrong,
K.~Berkelman,
K.~Cranmer,
D.P.S.~Ferguson,
Y.~Gao,$^{29}$
S.~Gonz\'{a}lez,
O.J.~Hayes,
H.~Hu,
S.~Jin,
J.~Kile,
P.A.~McNamara III,
J.~Nielsen,
Y.B.~Pan,
\mbox{J.H.~von~Wimmersperg-Toeller}, 
W.~Wiedenmann,
J.~Wu,
Sau~Lan~Wu,
X.~Wu,
G.~Zobernig
\nopagebreak
\begin{center}
\parbox{15.5cm}{\sl\samepage
Department of Physics, University of Wisconsin, Madison, WI 53706,
USA$^{11}$}
\end{center}\end{sloppypar}
\vspace{2mm}
\begin{sloppypar}
\noindent
G.~Dissertori
\nopagebreak
\begin{center}
\parbox{15.5cm}{\sl\samepage
Institute for Particle Physics, ETH H\"onggerberg, 8093 Z\"urich,
Switzerland.}
\end{center}\end{sloppypar}
}
\footnotetext[1]{Also at CERN, 1211 Geneva 23, Switzerland.}
\footnotetext[2]{Now at LAPP, 74019 Annecy-le-Vieux, France}
\footnotetext[3]{Also at Dipartimento di Fisica di Catania and INFN Sezione di
 Catania, 95129 Catania, Italy.}
\footnotetext[4]{Deceased.}
\footnotetext[5]{Also Istituto di Cosmo-Geofisica del C.N.R., Torino,
Italy.}
\footnotetext[6]{Now at Institut f\"ur Experimentelle Kernphysik, Universit\"at Karlsruhe, 76128 Karlsruhe, Germany.}
\footnotetext[7]{Supported by CICYT, Spain.}
\footnotetext[8]{Supported by the National Science Foundation of China.}
\footnotetext[9]{Supported by the Danish Natural Science Research Council.}
\footnotetext[10]{Supported by the UK Particle Physics and Astronomy Research
Council.}
\footnotetext[11]{Supported by the US Department of Energy, grant
DE-FG0295-ER40896.}
\footnotetext[12]{Now at Departement de Physique Corpusculaire, Universit\'e de
Gen\`eve, 1211 Gen\`eve 4, Switzerland.}
\footnotetext[13]{Supported by the Commission of the European Communities,
contract ERBFMBICT982874.}
\footnotetext[14]{Also at Rutherford Appleton Laboratory, Chilton, Didcot, UK.}
\footnotetext[15]{Permanent address: Universitat de Barcelona, 08208 Barcelona,
Spain.}
\footnotetext[16]{Supported by the Bundesministerium f\"ur Bildung,
Wissenschaft, Forschung und Technologie, Germany.}
\footnotetext[17]{Supported by the Direction des Sciences de la
Mati\`ere, C.E.A.}
\footnotetext[18]{Supported by the Austrian Ministry for Science and Transport.}
\footnotetext[19]{Now at SAP AG, 69185 Walldorf, Germany}
\footnotetext[20]{Now at Groupe d' Astroparticules de Montpellier, Universit\'e de Montpellier II, 34095 Montpellier, France.}
\footnotetext[21]{Now at D\'epartement de Physique, Facult\'e des Sciences de Tunis, 1060 Le Belv\'ed\`ere, Tunisia.}
\footnotetext[22]{Supported by the US Department of Energy,
grant DE-FG03-92ER40689.}
\footnotetext[23]{Now at Skyguide, Swissair Navigation Services, Geneva, Switzerland.}
\footnotetext[24]{Also at Dipartimento di Fisica e Tecnologie Relative, Universit\`a di Palermo, Palermo, Italy.}
\footnotetext[25]{Now at CERN, 1211 Geneva 23, Switzerland.}
\footnotetext[26]{Now at Honeywell, Phoenix AZ, U.S.A.}
\footnotetext[27]{Now at INFN Sezione di Roma II, Dipartimento di Fisica, Universit\`a di Roma Tor Vergata, 00133 Roma, Italy.}
\footnotetext[28]{Now at Centre de Physique des Particules de Marseille, Univ M\'editerran\'ee, F-13288 Marseille, France.}
\footnotetext[29]{Also at Department of Physics, Tsinghua University, Beijing, The People's Republic of China.}
\footnotetext[30]{Now at SLAC, Stanford, CA 94309, U.S.A.}
\footnotetext[31]{Now at LBNL, Berkeley, CA 94720, U.S.A.}
\setlength{\parskip}{\saveparskip}
\setlength{\textheight}{\savetextheight}
\setlength{\topmargin}{\savetopmargin}
\setlength{\textwidth}{\savetextwidth}
\setlength{\oddsidemargin}{\saveoddsidemargin}
\setlength{\topsep}{\savetopsep}
\normalsize
\newpage
\pagestyle{plain}
\setcounter{page}{1}

\section{Introduction}

In the minimal supersymmetric extensions of the Standard Model
(MSSM)~\cite{mssm}, it is usually assumed that $R$-parity,
$R_p=(-1)^{3B+L+2S}$, is conserved~\cite{rpc}. Here $B$ denotes the
baryon number, $L$ the lepton number and $S$ the spin of a field. The
conservation of $R$-parity is not required theoretically and models in
which $R$-parity is violated can be constructed which are compatible
with existing experimental constraints.

The $R$-parity violating terms of the superpotential~\cite{supot} 
include a lepton violating contribution $\lambda_{ijk} L_i L_j {\bar{E}}_k$, 
where $L_{i(j)}$ are the lepton doublet superfields, $\bar{E_k}$ is the lepton 
singlet superfield, and $\lambda_{ijk}$ is the Yukawa coupling corresponding 
to a particular choice of generational indices $i,j,k$. 
A nonzero value for $\lambda_{ijk}$ implies that the lightest
supersymmetric particle (LSP) is not stable and that sparticles
could be produced singly. In particular, single sneutrino production would be
possible in ${\rm e}^+{\rm e}^-$ collisions via the ${\rm e} \gamma \to 
\tilde{\nu}_j \ell_k$ subprocess.
For this process, the kinematic reach is typically twice that of pair 
production. As the production cross section depends on the value 
of the Yukawa coupling, an observation of supersymmetry in this channel 
would allow the Yukawa coupling to be directly measured. 

In this paper, the results of searches for single sneutrino production via 
the ${\rm e} \gamma \to \tilde{\nu}_j \ell_k$ subprocess in the data recorded 
by the ALEPH detector at LEP are reported, under the assumption of a single 
dominant $\lambda$ coupling and degenerate sneutrino masses. 
The paper is organised as follows.
In Section 2, the production and decay of single sneutrinos are discussed. 
In Section 3, the ALEPH detector is briefly described. The details of the data 
used, the simulation of the signal and the various Standard Model backgrounds
are given in Section 4.
In Section 5, the selections for the various final states are discussed. 
Finally, the results are given and the
corresponding limits on the $\lambda$ couplings are presented in Section 6.

\section{Single Sneutrino Production and Decay}

\begin{figure}[!t]
\begin{center}
\begin{picture}(160,40)
\put(10,0){\epsfxsize140mm\epsfbox{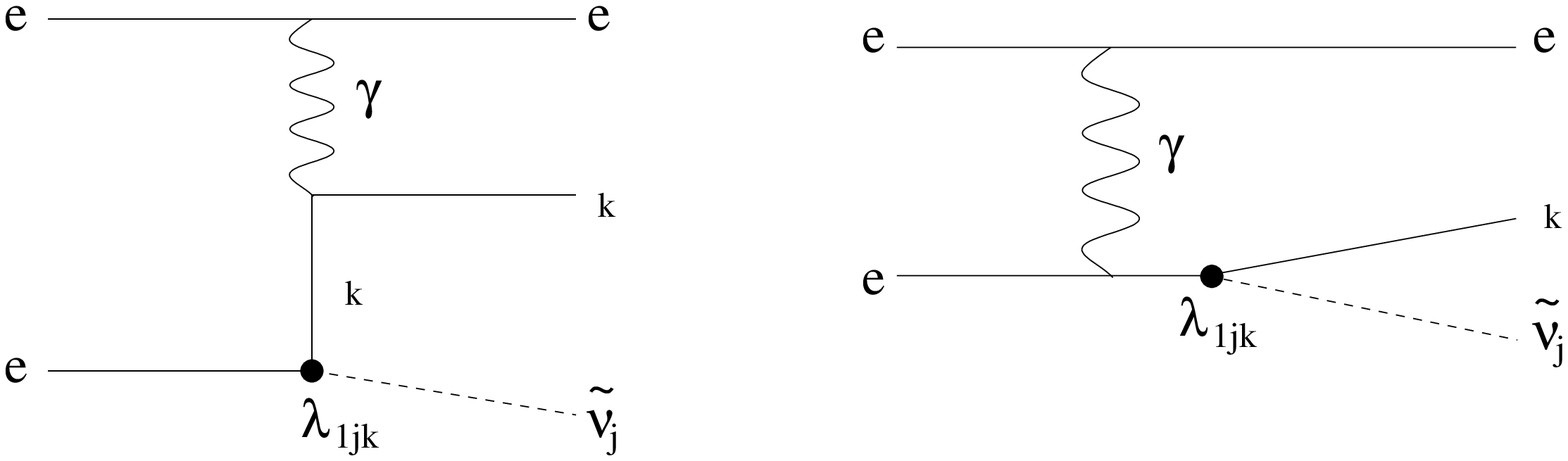}}
\put(38.6,15.5){$\ell$}
\put(61.2,23.5){$\ell$}
\put(145.7,22){$\ell$}
\end{picture}
\end{center}
\caption{\label{fig:prod}{\ninerm Production of a single sneutrino via the 
$\scriptstyle {\rm e} \gamma \to \tilde{\nu}_j\ell_k$ subprocess. 
The $\scriptstyle LL\bar{E}$ $\scriptstyle R$-parity violating vertex 
is indicated by the black dot.}} 

\begin{center}
\begin{picture}(160,35)
\put(10,0){\epsfxsize140mm\epsfbox{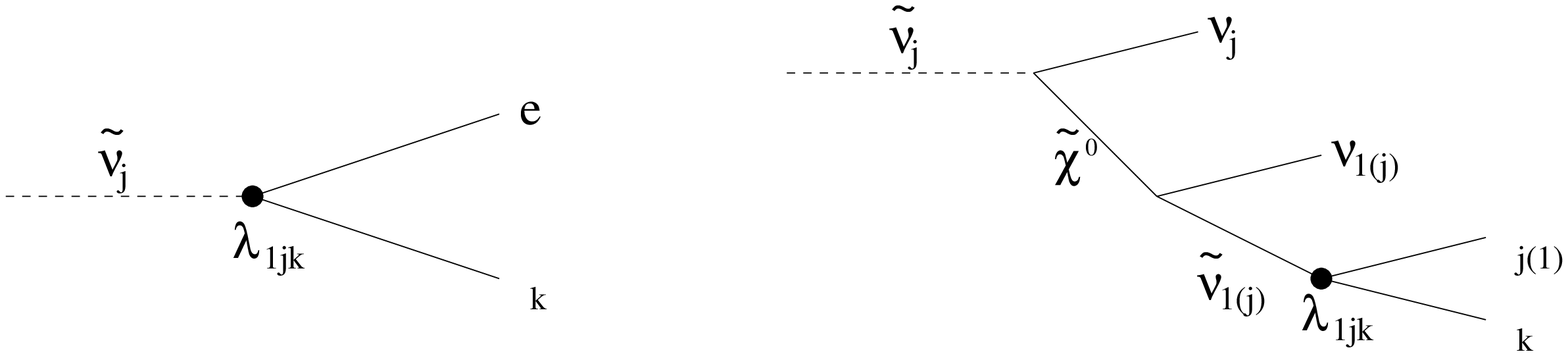}}
\put(54.8,5.5){$\ell$}
\put(142.3,8.9){$\ell$}
\put(142.3,2){$\ell$}
\end{picture}
\end{center}
\caption{\label{fig:decay}{\ninerm $\scriptstyle R$-parity violating 
sneutrino decays: (left) direct decay, (right) indirect decay via a 
neutralino. The $\scriptstyle LL\bar{E}$ $\scriptstyle R$-parity 
violating vertex is indicated by the black dot.}} 
\end{figure}

The tree-level Feynman diagrams for the production of a single sneutrino, 
via the ${\rm e} \gamma \to \tilde{\nu}_j\ell_k$ subprocess, in 
${\rm e}^+{\rm e}^-$ interactions are shown in Fig.~\ref{fig:prod}. 
As indicated in Fig.~\ref{fig:decay}, the sneutrino can decay {\it directly},
$\tilde{\nu}\to {\rm e}\ell$ or, if the neutralino mass is lower than that of 
the sneutrino, {\it indirectly} via the lightest neutralino, $\tilde{\nu}\to 
\nu\chi$. The possibility of indirect decays via higher mass neutralinos 
or charginos is not considered in this paper. For small $\lambda$ 
couplings, the indirect decays normally dominate over the direct decay, 
if accessible. The situation is reversed for large $\lambda$ couplings 
or small mass differences between the sneutrino and the neutralino. 
These two extreme configurations are addressed in turn in the following.

\begin{figure}[!h]
\begin{center}
\epsfig{file=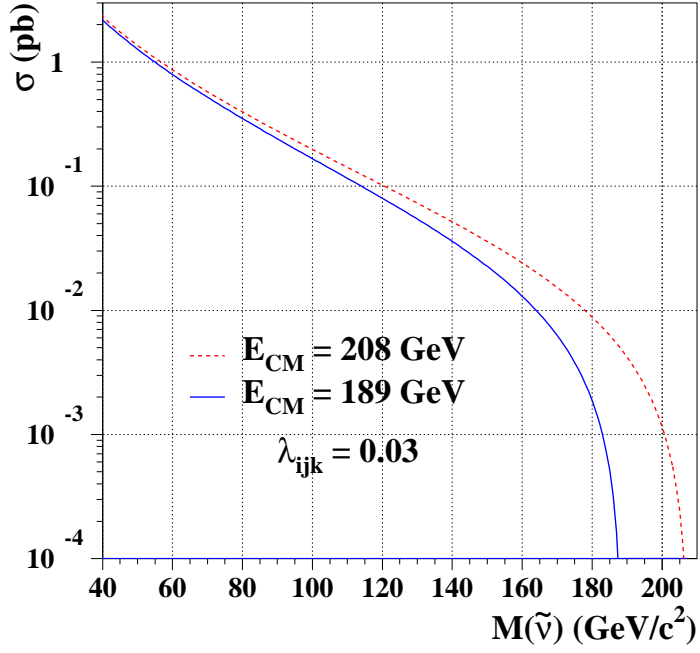,width=0.6\linewidth}
\vspace*{-0.5cm}
\caption{\label{fig:sectheo}{\ninerm The cross section for single sneutrino
production, assuming $\scriptstyle \lambda_{122(132)}=$ 0.03, as a function 
of the sneutrino mass. The full curve is the cross section at a centre-of-mass 
energy of 189\,GeV and the dashed curve is the cross section at a 
centre-of-mass energy of 208\,GeV. }}

\end{center}
\end{figure}

The production cross section~\cite{allanach} depends on the assumed value 
of $\lambda$, the mass of the sneutrino, the mass of the lepton produced 
in association with the sneutrino and the centre-of-mass energy. 
Figure~\ref{fig:sectheo} shows the cross section for ${\rm e}^+{\rm e}^- 
\to {\rm e} \mu {\tilde{\nu}}$ as a function of the sneutrino mass, for 
two different centre-of-mass energies, and assuming $\lambda_{122(132)}=0.03$.
The cross section for  ${\rm e}^+{\rm e}^- \to {\rm e} \tau {\tilde{\nu}}$ 
is about a factor two lower. In the calculation of the cross section, it 
is assumed that the deflection angle of the beam electron is less than 
30\,mrad, implying that the electron remains in the beam pipe and 
is not observed in the detector. 

With this production process a sneutrino can be produced via seven of the 
nine possible $\lambda$ couplings. Throughout this analysis, only one 
nonzero $\lambda$ coupling is assumed, thus the sneutrino is also assumed 
to decay via the same coupling. A summary of the possible final states 
corresponding to the seven accessible couplings is shown in 
Table~\ref{tab:tdecay}. All the direct and indirect decays, except those 
for the 121 and 131 couplings, are addressed in the following. For these 
two couplings the ALEPH search for resonant sneutrino production~\cite{rpv202} 
already provides stringent limits, which cannot be improved upon 
with the current analysis. For the decay of the neutralino,
it is assumed that the sneutrinos are degenerate in mass, leading to
two possible decays of equal probability.  

\vspace*{1cm}

\begin{table}[!h]
\caption{\ninerm The final states produced by the different sneutrino flavours
for the various couplings for the direct decays and the indirect decays to
the lightest neutralino. The last two leptons listed in the final 
state are the decay products of the sneutrino. The non detected beam 
electron is not included in the final state. The charge conjugate reaction, 
in which the beam positron radiates the photon, is also possible but not 
indicated. The entries marked with a dash are those for which
the sneutrino flavour cannot be produced by the coupling in question.
\newline}
\label{tab:tdecay}
\centering
\begin{tabular}{|c||c|c|c||c|c|c||l|}\hline
$\lambda$ & \multicolumn{3}{c||}{Direct Decays}&\multicolumn{4}{c|}{Indirect
Decays}\\ \cline{2-8}
ijk&$\tilde{\nu}_{\rm e}$&$\tilde{\nu}_\mu$&$\tilde{\nu}_\tau$&$\tilde{\nu}_{\rm e}$&$\tilde{\nu}_\mu$&$\tilde{\nu}_\tau$&\multicolumn{1}{c|}{$\chi$ Decay} \\
 \hline \hline
121 &$\mu^+{\rm e}^+\mu^-$  &${\rm e}^+{\rm e}^+{\rm e}^-$&-&$\mu^+\nu_{\rm e}\chi$&${\rm e}^+\nu_\mu\chi$&-&$\chi\rightarrow\nu_\mu {\rm e}^+{\rm e}^-\;or\;\nu_{\rm e} {\rm e}^\pm\mu^\mp$ \\ \hline

131 &$\tau^+{\rm e}^+\tau^-$&-&${\rm e}^+{\rm e}^+{\rm e}^-$&$\tau^+\nu_{\rm e}\chi$&-&${\rm e}^+\nu_\tau\chi$&$\chi\rightarrow\nu_\tau {\rm e}^+{\rm e}^-\;or\;\nu_{\rm e} {\rm e}^\pm\tau^\mp$\\ \hline

122 &-&$\mu^+{\rm e}^+\mu^-$ &-&-&$\mu^+\nu_\mu\chi$ &-&$\chi\rightarrow\nu_\mu {\rm e}^\pm\mu^\mp\;or\;\nu_{\rm e}\mu^+\mu^-$\\ \hline
132 &-&-&$\mu^+{\rm e}^+\mu^-$ &-&-&$\mu^+\nu_\tau\chi$ &$\chi\rightarrow\nu_\tau {\rm e}^\pm\mu^\mp\;or\;\nu_{\rm e}\mu^\pm\tau^\mp$\\ \hline
123 &-&$\tau^+{\rm e}^+\tau^-$ &-&-&$\tau^+\nu_\mu\chi$&-&$\chi\rightarrow\nu_\mu {\rm e}^\pm\tau^\mp\;or\;\nu_{\rm e} \mu^\pm\tau^\mp$\\ \hline
133 &-&-&$\tau^+{\rm e}^+\tau^-$&-&-&$\tau^+\nu_\tau\chi$&$\chi\rightarrow\nu_\tau {\rm e}^\pm\tau^\mp\;or\;\nu_{\rm e}\tau^+\tau^-$\\ \hline
231 &-&$\tau^+{\rm e}^+\tau^-$&$\mu^+{\rm e}^+\mu^-$&-&$\tau^+\nu_\mu\chi$&$\mu^+\nu_\tau\chi$&$\chi\rightarrow\nu_\tau {\rm e}^\pm\mu^\mp\;or\;\nu_\mu {\rm e}^\pm\tau^\mp$\\ \hline
\end{tabular}
\end{table}

\section{\label{aleph.detector}The ALEPH Detector}

The ALEPH detector is described in detail in Ref.~\cite{bib:detectorpaper}. 
An account of the performance of the detector and a description of the 
standard analysis algorithms can be found in Ref.~\cite{bib:performancepaper}.
Here, only a brief description of the detector components and the algorithms 
relevant for this analysis is given.

The trajectories of charged particles are measured with a silicon
vertex detector, a cylindrical drift chamber, and a large time
projection chamber (TPC). The central detectors are immersed in a
1.5\,T axial magnetic field provided by a superconducting solenoidal
coil.  The electromagnetic calorimeter (ECAL), placed between the TPC
and the coil, is a highly segmented sampling calorimeter which is used
to identify electrons and photons and to measure their energies.  The
luminosity monitors extend the calorimetric coverage down to 34\,mrad
from the beam axis.  The hadron calorimeter (HCAL) consists of the
iron return yoke of the magnet instrumented with streamer tubes. It
provides a measurement of hadronic energy and, together with the
external muon chambers, muon identification.  The calorimetric and
tracking information are combined in an energy flow algorithm which
provides a list of charged and neutral objects. It also gives a measure 
of the total visible energy, and therefore the missing energy.

Electron identification is primarily based upon the matching between
the measured momentum of the charged track and the energy deposited in
the ECAL. Additional information from the shower profile in the ECAL
is also used. To take into account  energy loss due to bremsstrahlung and/or 
final state radiation photons, the sum of the ECAL energy deposits 
within a $5^\circ$ cone centred on the impact point of an isolated electron
is added to the energy determined by the tracking, if the additional energy 
is greater than 300\,MeV.

Muons are separated from hadrons by their characteristic pattern in the 
HCAL and the presence of associated hits in the muon chambers. A  
correction for final state radiation photons, similar to that used for the 
electrons, is also applied. In this case, the additional ECAL energy is only 
used if greater than 2\,GeV.

\section{Data and Monte Carlo Samples}

This analysis was performed using data collected with the ALEPH detector
during 1998-2001. It corresponds to an integrated luminosity of 
637.1\,$\mathrm{pb}^{-1}$ at centre-of-mass energies from 189 up to 209
GeV (Table~\ref{tab:luminosity}).
\begin{table}[!t]
\caption{\ninerm Integrated luminosity for all data collected by ALEPH 
in 1998-2001.\newline}
\label{tab:luminosity}
\centering
\begin{tabular}{|l||c|c|c|c|c|}\hline
$\sqrt{s}$   & 189\,GeV   &   192\,GeV   & 196\,GeV  &  200-202\,GeV & 203-209\,GeV \\ \hline
Luminosity 
& 174.2\,$\mathrm{pb}^{-1}$
&  28.9\,$\mathrm{pb}^{-1}$
&  78.8\,$\mathrm{pb}^{-1}$
& 128.2\,$\mathrm{pb}^{-1}$
& 227.0\,$\mathrm{pb}^{-1}$ \\ \hline
\end{tabular}
\end{table}

The signal was simulated with SUSYGEN~\cite{susygen}, modified to include
single sneutrino production. Final state radiation was implemented using
PHOTOS~\cite{photos} and tau decays via TAUOLA~\cite{tauola}. 
The events were passed through the full simulation chain of the ALEPH 
detector and the reconstruction program applied to the data. 

Monte Carlo samples simulating all relevant Standard Model processes were used,
corresponding to at least 100 times the collected luminosity in the data, 
except for the process $\gamma\gamma\rightarrow$ leptons where 10 million 
events were generated, corresponding to more than six times the collected 
luminosity in the data. Events from ${\rm e}^+{\rm e}^-\rightarrow 
{\rm q\bar q}$ at 189\,GeV and four-fermion events from ${\rm W}{\rm e}\nu$, 
ZZ and Zee were produced with PYTHIA~\cite{jetset}. The process 
${\rm e}^+{\rm e}^-\rightarrow {\rm q\bar q}$  above 189\,GeV was 
produced with KORALZ~\cite{koralz}. Pairs of W bosons were generated with 
KORALW~\cite{koralw}. Pair production of leptons was simulated 
with BHWIDE~\cite{bhwide} (electrons) and KORALZ~\cite{koralz} 
(muons and taus), and the two-photon processes with PHOT02~\cite{phot02}.

\section{Event Selections}

The signal is characterised by a charged lepton, the decay products of the 
sneutrino and missing energy from an electron lost in the beam pipe. 
The flavours of the three final state leptons depend on the coupling involved.
In the case of indirect decays and decays involving taus, there can be 
significant additional missing momentum, of no particular preferred direction, 
due to the presence of neutrinos in the final state. 

A number of selections were developed to select the final states 
listed in Table~\ref{tab:tdecay} from the Standard Model backgrounds. 
As sneutrinos with masses less than $82\,\gevcc$ for a $\lambda > 10^{-7}$ 
are already excluded by the pair production search \cite{rpv202},  
the selections were optimised to give the minimum expected
$95\%$ C.L. excluded cross section~\cite{n95} for sneutrino masses above 
$80\,\gevcc$. For the indirect decays, mass differences between 
the sneutrino and neutralino less than $5\,\gevcc$ were not considered.
Neutralino masses less than $20\,\gevcc$ were not considered either
as they are already excluded~\cite{LLE}. 

A preselection requiring low charged multiplicity ($N_{\rm ch}<6$), 
at least one identified electron or muon ($N_{\rm e}+N_\mu>0$), and only 
a small amount of energy within a $12^\circ$ cone around the beam axis 
($E_{12}<20$\,GeV), allows most of the hadronic and semileptonic 
backgrounds to be rejected. The remaining backgrounds are low multiplicity 
leptonic events. 

The acoplanarity angle between jets (acopJet), obtained by forcing the 
event to form two jets using the Durham algorithm \cite{durham}, is a 
powerful tool to reject many of the backgrounds. The Zee and ZZ backgrounds 
are reduced using an anti-Z mass cut. Various requirements on the angles 
between the leptons and their angle with respect to the beam axis are 
also applied. In some cases, sliding cuts are adopted which depend upon 
the assumed sneutrino mass.   

For the direct decays to electron and muon ($\lambda_{122}$, $\lambda_{132}$ 
and $\lambda_{231}$), the invariant mass of the sneutrino can be 
reconstructed, unlike for the indirect decays and direct decays
with a tau in the final state. For these decays, requirements on 
the fraction of visible energy ($E_{\rm vis}/\sqrt{s}$), 
the direction of the missing momentum ($\theta_{\rm miss}$) 
and the amount of missing transverse momentum ($P_{\rm T}$) are applied. 

The various selections are outlined below and summarized in the Tables of the 
Appendix.

\subsection{Direct Decay via $\bf \lambda_{122}$ or $\bf \lambda_{132}$}

The direct sneutrino decays via $\lambda_{122}$ and $\lambda_{132}$ 
differ only in the flavour of the intermediate sneutrino produced.  
The decay products from the sneutrino (${\rm e}^+ \mu^-$) and the complete 
final state (${\rm e}^+ \mu^- \mu^+$) are identical for both couplings. 

The complete set of selection cuts is listed in 
Table~A1.
The selection requires one and only one identified electron in the event 
($N_{\rm e}$), and that the angle $\theta_{\rm e}$ of this electron with 
respect to the beam axis be large. In addition, it is required that the 
visible energy fraction be significant and that the total transverse 
momentum of the event be small. 

\begin{figure}
\vspace{-0.7cm}
\begin{center}
\epsfig{file=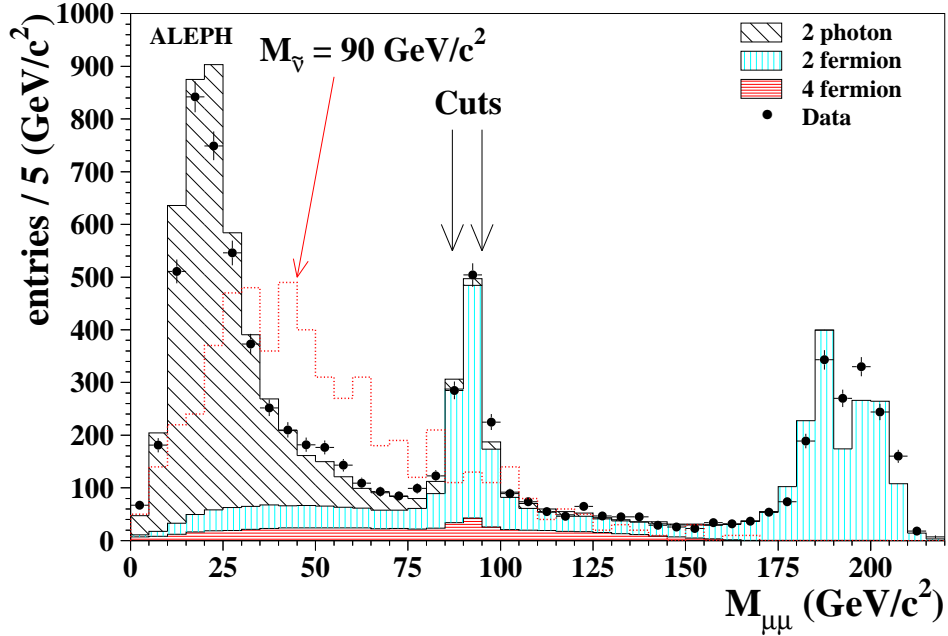,height=0.4\textheight}
\vspace{-0.5cm}
\caption{\label{fig:massmumu}{The $\mu^+ \mu^-$ 
invariant mass distribution, obtained after applying the preselection, for 
$\sqrt{s} = $ 189--209\,GeV: data (dots with error bars), expected 
background (solid histogram). The expected distribution (arbitrary 
normalisation) for a signal of $M_{\mathrm{\tilde{\nu}}} = $ 90\,\gevcc\
is also indicated (dotted histogram).}} 
\end{center}
\end{figure}

As in about 20\% of the cases a second lepton (in addition to the beam
electron) is outside the tracking acceptance, the selection treats the case 
of two or three charged particles separately. For the two track case, it 
is required that the other track be an identified muon of opposite charge 
to the electron and that the electron and muon be not back-to-back 
($\theta_{{\rm e}\mu}$). For the three track case, at least one muon
has to be identified. In addition, the muon and the non-electron track 
must not be back-to-back ($\theta_{\mu{\rm t}}$), and the invariant mass 
$M_{\mu{\rm t}}$ of these two  tracks must be inconsistent with the Z mass. 
Figure~\ref{fig:massmumu} shows the ${\rm Z} \rightarrow \mu^+ \mu^-$
mass peak obtained in data compared with Monte Carlo expectations, 
after applying the preselection.

\begin{figure}
\vspace{-1cm}
\begin{center}
\epsfig{file=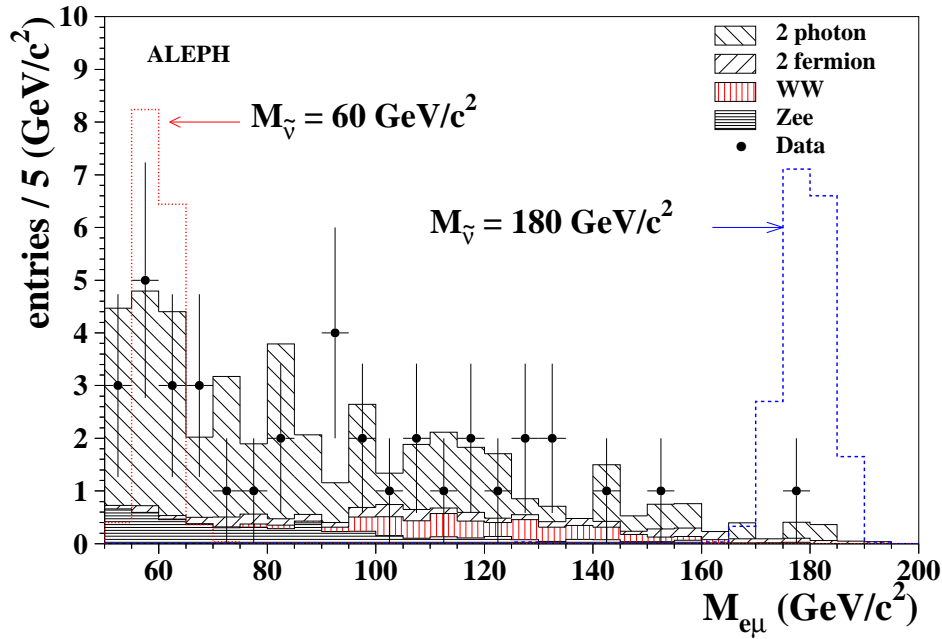,height=0.4\textheight}
\vspace{-0.5cm}
\caption{\label{fig:massemu}{  The ${\rm e}^+ \mu^-$ invariant mass 
distribution, obtained after applying the $\lambda_{1j2}$ selection, 
for $\sqrt{s} = $ 189--209\,GeV: data (dots with error bars), expected 
background (solid histogram). The expected distribution ($\lambda$~=~0.03) 
for a signal of $M_{\mathrm{\tilde{\nu}}}=60\,\gevcc$ (dotted histogram) and 
$M_{\mathrm{\tilde{\nu}}}=180\,\gevcc$ are also indicated (dashed 
histogram).}} 
\end{center}
\end{figure}

The efficiency of the complete selection increases from 25\% for a sneutrino
mass of $60\,\gevcc$ to 55\% for a sneutrino mass of $180\,\gevcc$. 
Summing over all centre-of-mass energies a total of 38 events is observed 
in the data while 46.7 are expected from Standard Model backgrounds. 
Figure~\ref{fig:massemu} shows the ${\rm e}^+\mu^-$ mass distribution of the 
selected events. As indicated in the figure, a signal would produce a peak 
centred around the mass of the sneutrino. For the extraction of the 
$\lambda$ exclusions , events are counted within a mass window of variable 
width which is scanned across this ${\rm e}^+ \mu^-$ mass distribution.

\subsection{\label{sec-direct1j3} Direct Decay via $\bf \lambda_{123}$ or
$\bf \lambda_{133}$}

The direct sneutrino decays via $\lambda_{123}$ and $\lambda_{133}$ 
differ only in the flavour of the intermediate sneutrino produced.  
The decay products from the sneutrino (${\rm e}^+ \tau^-$) and the complete 
final state (${\rm e}^+ \tau^- \tau^+$) are identical for both couplings.

The complete set of selection cuts is listed in 
Table~A2.
For the two tau's, the 1/1 and 1/3 prong decays are considered, leading to 
a final state with three or five charged particles. The tau decays into 
electrons and muons are considered in the requirements on the numbers of 
electrons and muons in the event. As the events contain neutrinos from the 
tau decays, a large transverse momentum and significant missing energy 
are required . The missing momentum is also required not to be in the 
direction of the beam axis.

The most energetic electron in the event, usually produced directly from the 
sneutrino decay, is required to have an energy $E_{\rm emax}$ greater than 
20\,GeV and the direction of this electron $\theta_{\rm emax}$ 
is required to be away from the beam axis. 

The assignment of the charged and neutral objects to the two taus is performed 
by removing the most energetic electron from the event. Requirements are made 
that the tau ``jets'' have opposite charge and be not back-to-back 
($\theta_{\tau_1 \tau_2}$). Additional requirements on the transverse 
momentum of the most energetic tau jet ($P_{\rm T(maxEjet)}$) and the angle
between the most energetic electron and the same sign tau jet  
($\theta_{{\rm emax}^\pm/\tau^\pm}$) are also applied. 

\begin{figure}
\vspace{-1cm}
\begin{center}
\epsfig{file=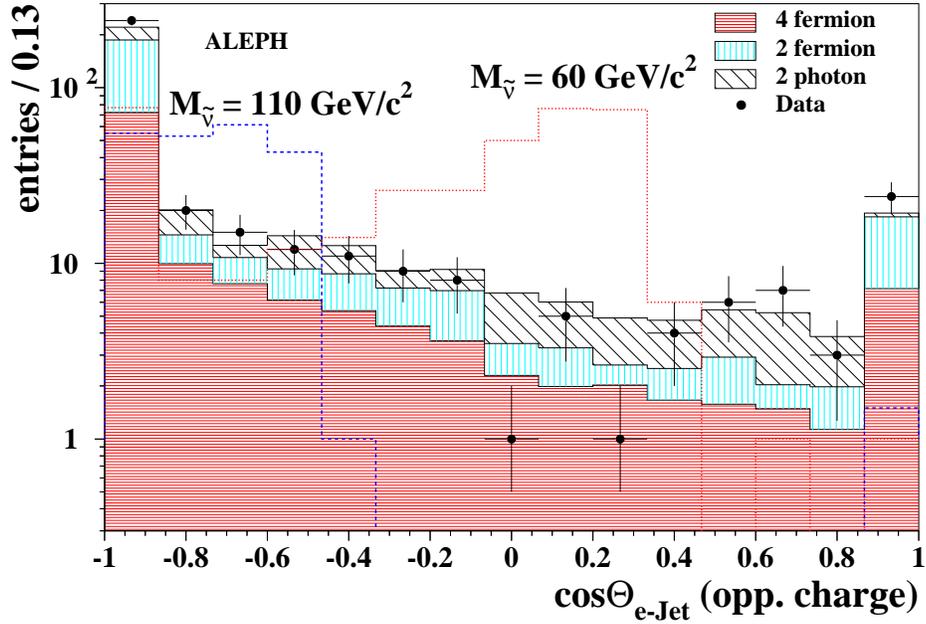,height=0.4\textheight}
\vspace{-0.5cm}
\caption{\label{fig:cosejop}{  The distribution, at preselection level, 
of the angle between the most energetic electron 
and the opposite-sign tau jet, used in the $\lambda_{1j3}$ selection, for $\sqrt{s} = $ 189--209\,GeV: 
data (dots with error bars), expected background (solid histogram).
The expected distribution (arbitrary normalisation) 
for a signal of $M_{\mathrm{\tilde{\nu}}}=60\,\gevcc$ (dotted histogram) and
$M_{\mathrm{\tilde{\nu}}}=160\,\gevcc$ are also indicated (dashed histogram).}} 
\end{center}
\end{figure}

After all selection criteria, the efficiency for the signal rises from 
15\% for a sneutrino mass of $60\,\gevcc$ to 35\% for a $160\,\gevcc$ 
sneutrino. A total of 13 events is selected in the data, consistent with 
the 17.1 events expected from the Standard Model backgrounds. 
Due to the presence of neutrinos in its decay, the mass of the sneutrino 
cannot be directly reconstructed from the decay products. Nevertheless, 
the energy of the most energetic electron and the angle between this electron 
and the opposite-sign tau ($\theta_{{\rm emax}^\pm/\tau^\mp}$) are sensitive 
to the sneutrino mass. These variables are therefore used to establish 
sliding cuts for extraction of the $\lambda$ exclusions. 
Figure \ref{fig:cosejop} shows the distribution of 
$\theta_{\mathrm{emax^\pm/\tau^\mp}}$ in the data, after
the preselection, compared to Monte Carlo expectations.
       
\subsection {Direct Decay via $\bf \lambda_{231}$}

The direct decay via the $\lambda_{231}$ coupling can proceed via two 
different flavours of sneutrino. The production of a $\tilde{\nu}_{\tau}$ 
leads to a final state with two muons and an electron. The decay via a 
$\tilde{\nu}_{\mu}$ leads to a final state with two taus and an electron. 
Due to the mass of the tau produced in association with the 
$\tilde{\nu}_{\mu}$, this channel is suppressed by a factor 2.25 with respect
to the $\tilde{\nu}_{\tau}$ channel.

As the final states are identical to those for the direct decays via the 
$\lambda_{1j2}$ and $\lambda_{1j3}$ couplings, previously discussed, no 
additional selection was developed for this coupling.   

\subsection{Indirect Decay via $\bf \lambda_{122}$}

Depending on the neutralino decay, the indirect decay via the $\lambda_{122}$ 
coupling can produce two distinct final states of equal probability, 
either two muons and an electron plus neutrinos or three muons plus neutrinos. 

\begin{figure}
\vspace{-0.7cm}
\begin{center}
\epsfig{file=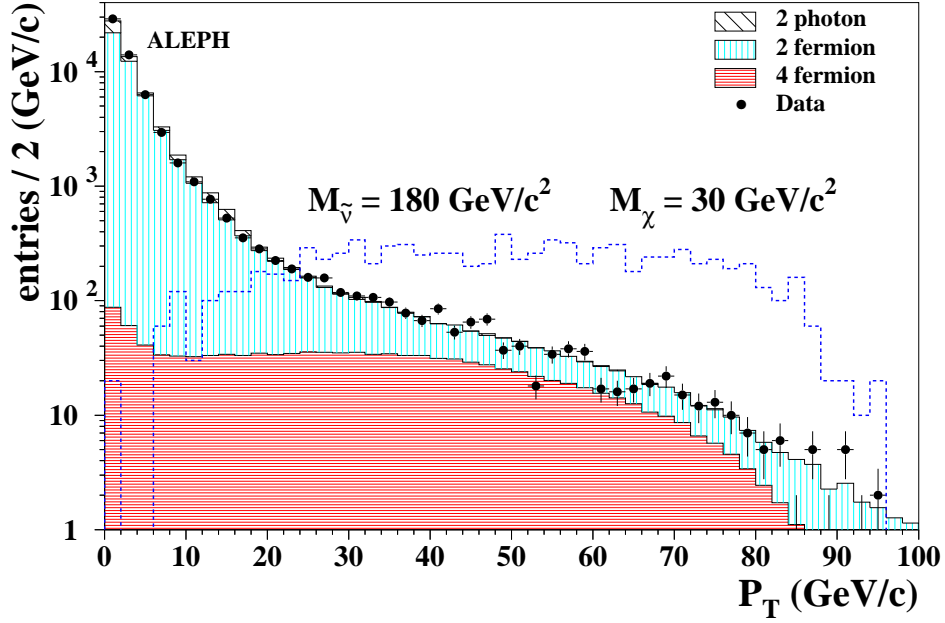,height=0.4\textheight}
\vspace{-0.5cm}
\caption{\label{fig:pt}{  The distribution of the missing transverse momentum, at preselection 
level, for $\sqrt{s} = $ 189--209\,GeV: data (dots with error bars), expected background (solid histogram).
The expected distribution for a signal (arbitrary normalisation)
with $M_{\mathrm{\tilde{\nu}}}=180\,\gevcc$ and 
$M_{\mathrm{\chi}}=30\,\gevcc$ is also indicated (dashed histogram).}} 
\end{center}
\end{figure}

For the selection 
(Table~A3), 
three charged tracks are 
required of which at least two must be identified as muons. Cuts requiring 
a large transverse momentum in the event and a large missing momentum  
not pointing along the beam axis are used. Figure~\ref{fig:pt} shows the 
distrbution of the missing $P_{\rm T}$ at the preselection level.
For the case of three identified muons, a cut removing combinations 
consistent with the Z mass, is used to reject the ZZ background. For the case 
of two identified muons an additional cut requiring the polar angle of the 
non-muon track ($\theta_{\rm t}$) to be away from the beam axis is applied.  

The efficiency of the selection for the signal varies between 40\% and 50\% 
and is insensitive to the mass of the neutralino involved in the decay. 
No event is selected in the data, consistent with the Standard Model 
background expectation of 2.1 events.  

\subsection{Indirect Decay via $\bf \lambda_{132}$}

Depending on the neutralino decay, the indirect decay via the $\lambda_{132}$ 
coupling can produce two distinct final states of equal probability,  
either two muons and an electron plus neutrinos or two muons and a tau plus 
neutrinos. 

For the selection 
(Table~A4) 
only the one-prong decays of 
the tau are considered. Three charged tracks are therefore required, of 
which at least one must be identified as a muon. Only one identified 
electron is allowed in the event. Cuts requiring a large transverse 
momentum and a large missing energy whose direction is away from the beam 
axis are also applied. A cut on the acoplanarity of two jets is also used. 

The efficiency of the selection for the signal is 30\%--40\% depending on 
the sneutrino  mass and is insensitive to the mass of the neutralino 
involved in the decay. Three events are selected in the data, consistent 
with the Standard Model background expectation of 2.2 events.  

\subsection{Indirect Decay via $\bf \lambda_{123}$}

Depending on the neutralino decay, the indirect decay via the $\lambda_{123}$ 
coupling can produce two distinct final states of equal probability,  
either two taus with an electron or two taus with a muon. 

The complete list of selection criteria can be found in 
Table~A5. 
A total of three or five charged particles 
is required in the event, corresponding to the case where the two taus 
decay into 1/1 prong or 1/3 prongs. Due to the presence of neutrinos 
in each of the sneutrino, neutralino and tau decays, this channel is 
characterised by a small visible energy and large transverse momentum.

\begin{figure}
\vspace{-1.1cm}
\begin{center}
\epsfig{file=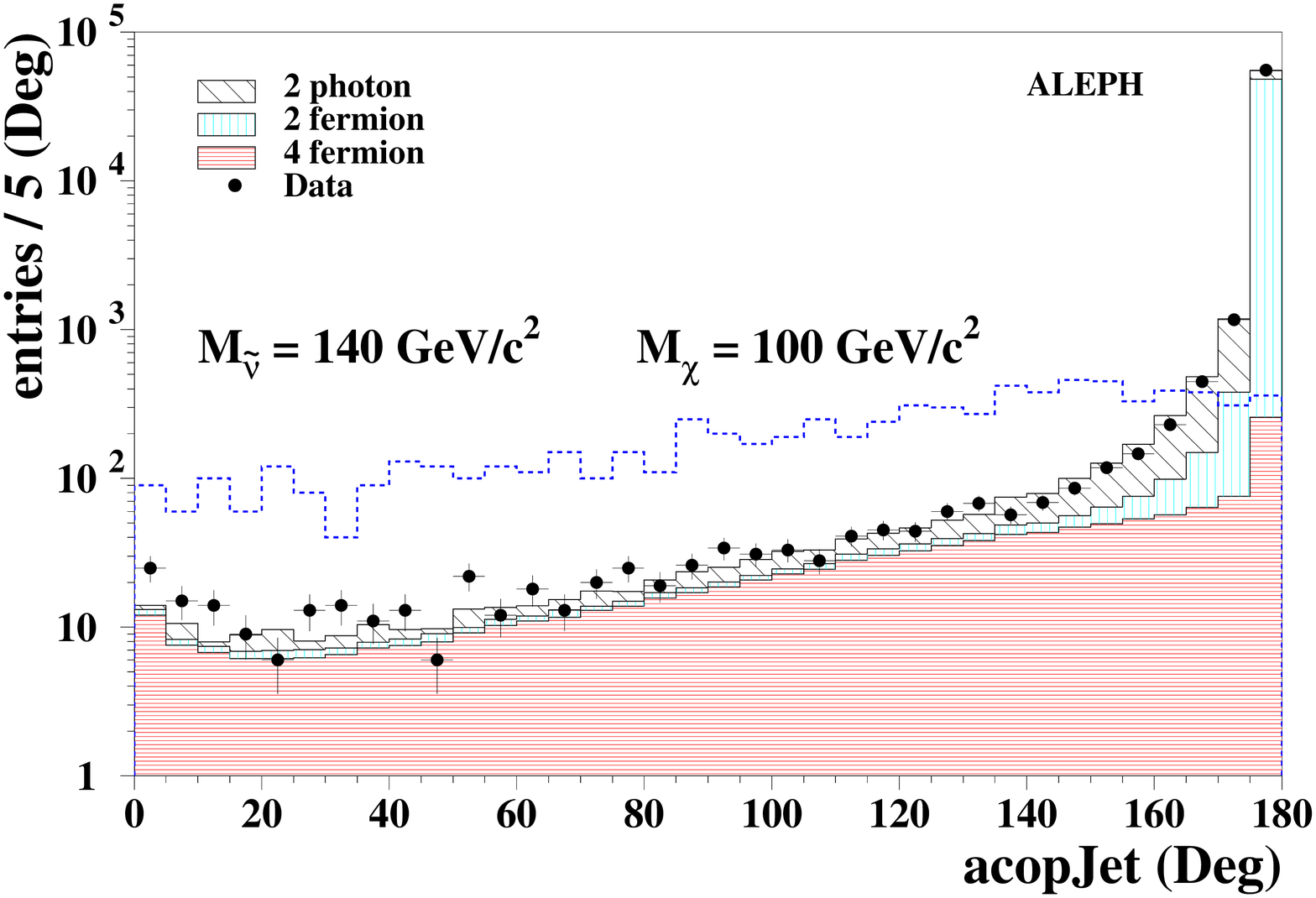,height=0.4\textheight}
\vspace{-0.5cm}
\caption{\label{fig:acopjet}{  The distribution of the acoplanarity between two jets, 
at the preselection level,  
for $\sqrt{s} = $ 189--209\,GeV: data (dots with error bars), expected background (solid histogram).
The expected distribution (arbitrary normalisation) 
for a signal of $M_{\mathrm{\tilde{\nu}}}=140\,\gevcc$ 
and $M_{\mathrm{\chi}}=100\,\gevcc$ (dashed histogram) is also indicated.}} 
\end{center}
\end{figure}

Cuts on the acoplanarity angle (acopJet) between the two jets, the transverse 
momentum ($P_{\rm T(maxEjet)}$) of the most energetic jet , and the angle 
$\theta_{{\rm jet}_1,{\rm jet}_2}$ between the two jets are effective at
reducing the backgrounds from $\gamma \gamma \rightarrow \tau^+\tau^-$ and 
${\rm Z} \to \tau^+\tau^-$. Figure~\ref{fig:acopjet} shows the distribution 
of acopJet at the preselection level. Finally a cut on $y_{23}$, the Durham 
distance for the transition from two to three jets, is applied to reject 
$\gamma \gamma \rightarrow \ell^+\ell^-$ and  ${\rm e}^+{\rm e}^- \rightarrow 
\ell^+\ell^-$. 

The efficiency of this selection for the signal rises from 25\% for a 
sneutrino mass of $60\,\gevcc$ to 40\% for a sneutrino mass of $180\,\gevcc$.
The dependence of the efficiency on the neutralino mass is small. 
A total of 14 events is selected in the data, consistent with the 14.9 events 
expected from the Standard Model backgrounds. 

\subsection{Indirect Decay via $\bf \lambda_{133}$}

Depending on the neutralino decay, the indirect decay via the $\lambda_{133}$ 
coupling can produce two distinct final states of equal probability, either 
two taus and an electron plus neutrinos or three taus plus neutrinos. 
These events are therefore characterised by a large number of taus in the 
final state.

The selection is summarised in 
Table~A6. 
Three or five charged tracks are demanded, 
implying that the 1/1 and 1/3 prong decays of the 
two taus are considered in the two-tau final state (97.5\% of the cases), 
and the 1/1/1 and 1/1/3 prong decays are considered for the three-tau final 
state (93.5\% of the cases).  Up to two identified electrons or muons are 
allowed in the event. Due to the presence of neutrinos, cuts requiring a 
large transverse momentum and a large missing momentum not pointing along 
the beam axis are applied. A cut on the acoplanarity of the event is also 
used. For the case of three charged tracks the backgrounds are larger than 
the five-track case, necessitating tighter cuts.

The efficiency of the selection for the signal is 20\%--35\%, depending on 
the sneutrino mass and is insensitive to the mass of the neutralino involved 
in the decay. A total of 16 events is selected in the data, consistent with 
the Standard Model background expectation of 16.6 events.  

\subsection{Indirect Decay via $\bf \lambda_{231}$}

The production and indirect decay of sneutrinos via the $\lambda_{231}$ 
coupling lead to different final states depending on whether a 
$\tilde{\nu}_\mu$ or a $\tilde{\nu}_\tau$ is produced and on the 
subsequent decay of the neutralino. Taking into account the different 
production cross sections of the two sneutrino flavours, the relative 
probabilities of each final state are 15.4\% $\tau \tau {\rm e}$, 
34.6\% $\tau \mu {\rm e}$ and 50\% $\mu \mu {\rm e}$.
  
The selection 
(Table~A7) 
treats the final states globally 
and only considers the one-prong decays of the tau. The simultaneous presence 
of an electron and a muon ($\sim 85\%$ of the cases) in the event, associated 
with a large amount of missing energy and acoplanarity, allows most of the 
backgrounds to be rejected.

The efficiency of the selection varies between 25\% and 45\%, depending on the 
sneutrino mass, and is insensitive to the neutralino mass involved in the 
decay. A total of nine events is selected in the data, while 10.1 are 
predicted from Standard Model backgrounds. 

\section{Results and Upper Limits on the $\bf \lambda$ Couplings}

The numbers of observed and expected events, obtained after applying the 
various selections (except the sliding cuts), are summarised in 
Table~\ref{tab:final_stat}. The observed numbers are in good agreement with 
the Standard Model expectations, indicating no evidence for single sneutrino 
production. Upper limits on the values of  the various $LL\bar{E}$ couplings 
are therefore derived. 

\begin{table}[!t]
\caption{\ninerm The selection efficiency, the numbers of expected background 
and the numbers of observed events for each selection (any sliding cuts 
removed) summed over all centre-of-mass energies.
\newline}
\label{tab:final_stat}
\centering
\begin{tabular}{|c||c|c|c|}\hline
Selection    & Efficiency (\%)        & Expected Background & Observed Data \\ \hline
$1j2$ direct   &  $25 \rightarrow 55$   & 46.7                & 38            \\ 
$1j3$ direct   &  $15 \rightarrow 35$   & 17.1                & 13            \\ 
123 indirect &  $25 \rightarrow 40$   & 14.9                & 14            \\ 
122 indirect &  $40 \rightarrow 50$   & 2.1                 & 0             \\ 
132 indirect &  $30 \rightarrow 40$   & 2.2                 & 3             \\ 
231 indirect &  $25 \rightarrow 45$   & 10.1                & 9             \\ 
133 indirect &  $20 \rightarrow 35$   & 16.6                & 16            \\ 
\hline
\end{tabular}
\end{table}
  
In the calculation of the limits, background subtraction was performed for the 
two- and four-fermion final states. The uncertainties on the background 
estimates were taken into account by reducing the amount of background 
subtracted. For the two-fermion processes, the amount of background 
subtracted was reduced by its Monte Carlo statistical uncertainty.   
The backgrounds subtracted for the four-fermion processes were reduced by
$20\%$ of their estimate. The $\gamma\gamma \rightarrow {\rm f\bar f}$ 
background was not subtracted.

The systematic uncertainties on the selection efficiencies are
4--5$\%$, dominated by the signal Monte Carlo statistical uncertainty, 
with small additional contributions from the simulation of lepton 
identification and energy flow reconstruction. These uncertainties 
were conservatively taken into account by reducing the selection 
efficiencies by one standard deviation.  

Possible differences between the data and the simulation 
in the distribution of the $E_{12}$ variable, due to the presence 
of beam-related backgrounds in the data, were investigated using events 
triggered at random beam crossing. The difference between data and 
Monte Carlo in the fraction of energy deposits in the forward detectors, 
when this energy is less than 20\,GeV, was found to be less than 5 per mill.

\begin{figure}
\vspace{-1cm}
\begin{center}
\epsfig{file=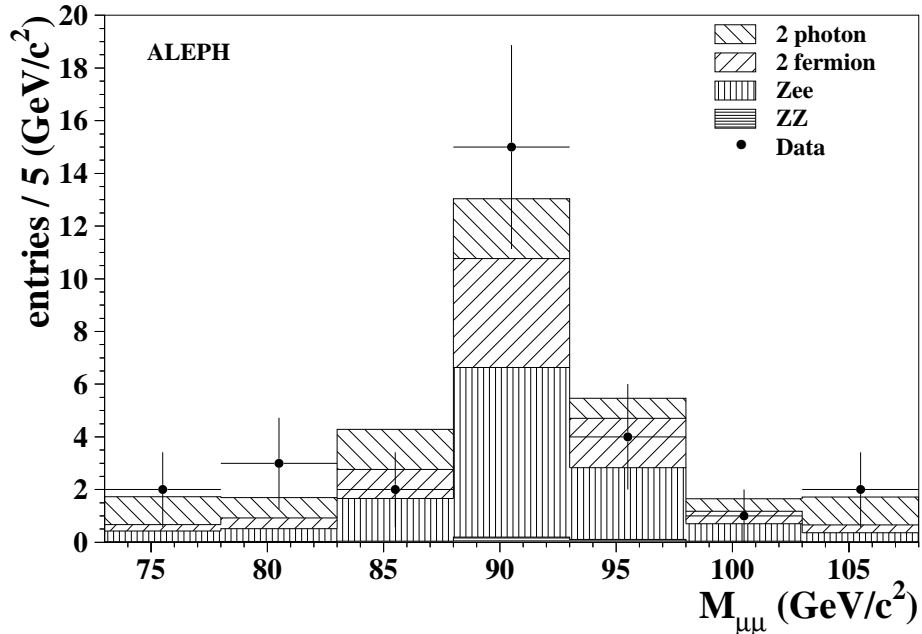,height=0.4\textheight}
\vspace{-0.5cm}
\caption{\label{fig:zeecheck}{  The invariant $\mu^+ \mu^-$ mass distribution, obtained 
using a modified $\lambda_{1j2}$ selection, for $\sqrt{s} = $ 189--209\,GeV: data (dots with error bars), 
expected background (solid histogram).}} 
\end{center}
\end{figure}

The Zee background, with ${\rm Z} \to \mu^+ \mu^-$, proceeds via a similar 
production mechanism and has the same final state as a sneutrino with a mass 
equal to the Z decaying directly through a $\lambda_{1j2}$ coupling, 
except that the electron in the sneutrino decay is replaced by a muon. 
As a check of the analysis procedure, the $\lambda_{1j2}$ selection was 
therefore modified to select the Zee background. Figure~\ref{fig:zeecheck} 
shows the $\mu^+ \mu^-$ mass spectrum after applying the modified selection. 
A clear peak around the Z mass is observed, consistent with 
expectations from the Zee and the two-fermion $\mu\mu$ simulations.

\begin{figure}[!t]
\vspace{-0.7cm}
\begin{center}
\epsfig{file=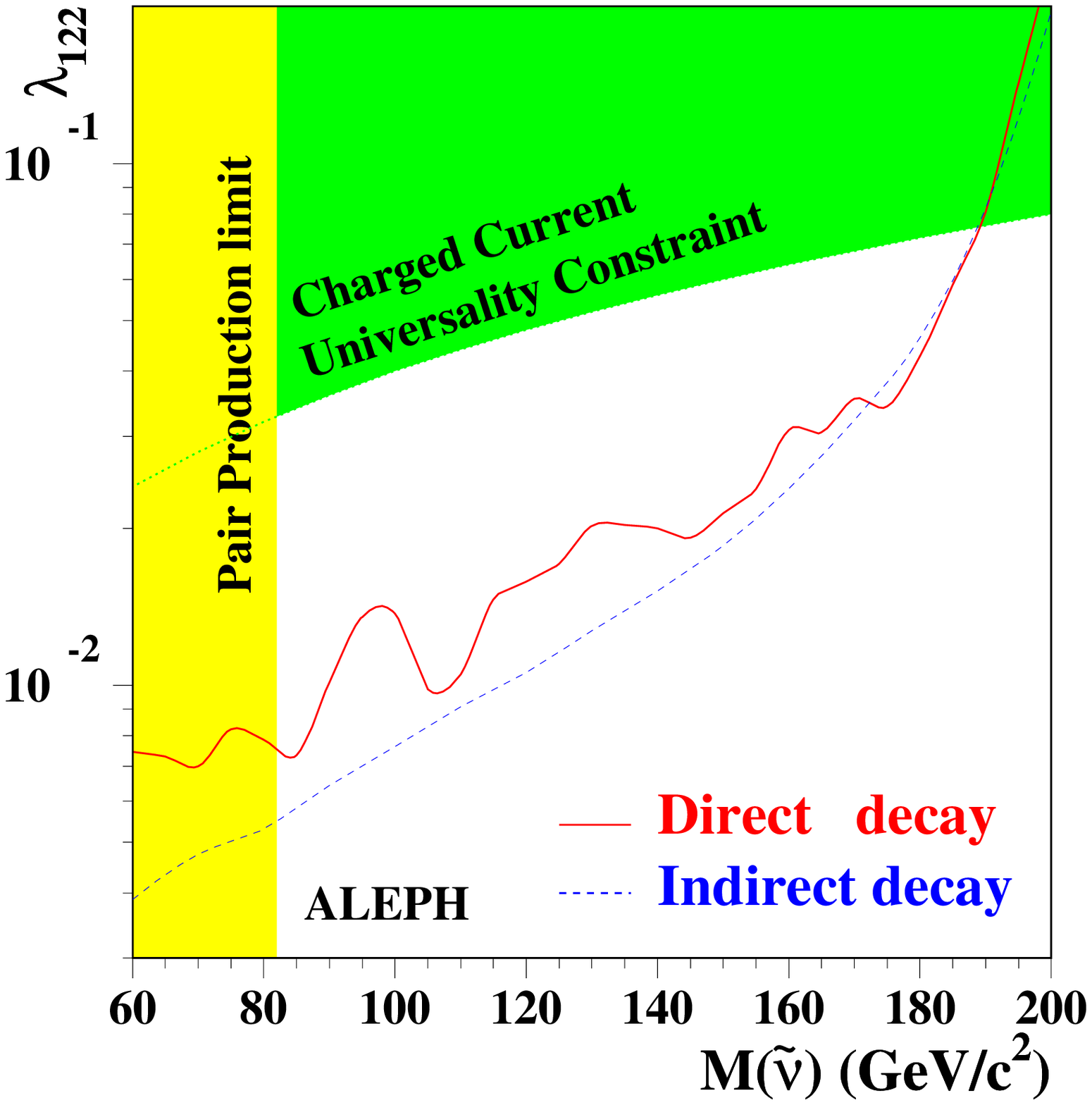,height=0.7\textheight}
\caption{\label{fig:122_excl}{  Observed 95\% CL upper limits on the 
$\lambda_{122}$ coupling for the direct and indirect decays.
The exclusions from the low energy bounds and the pair production
search are indicated.}} 
\end{center}
\end{figure}

\begin{figure}[!t]
\vspace{-1cm}
\begin{center}
\epsfig{file=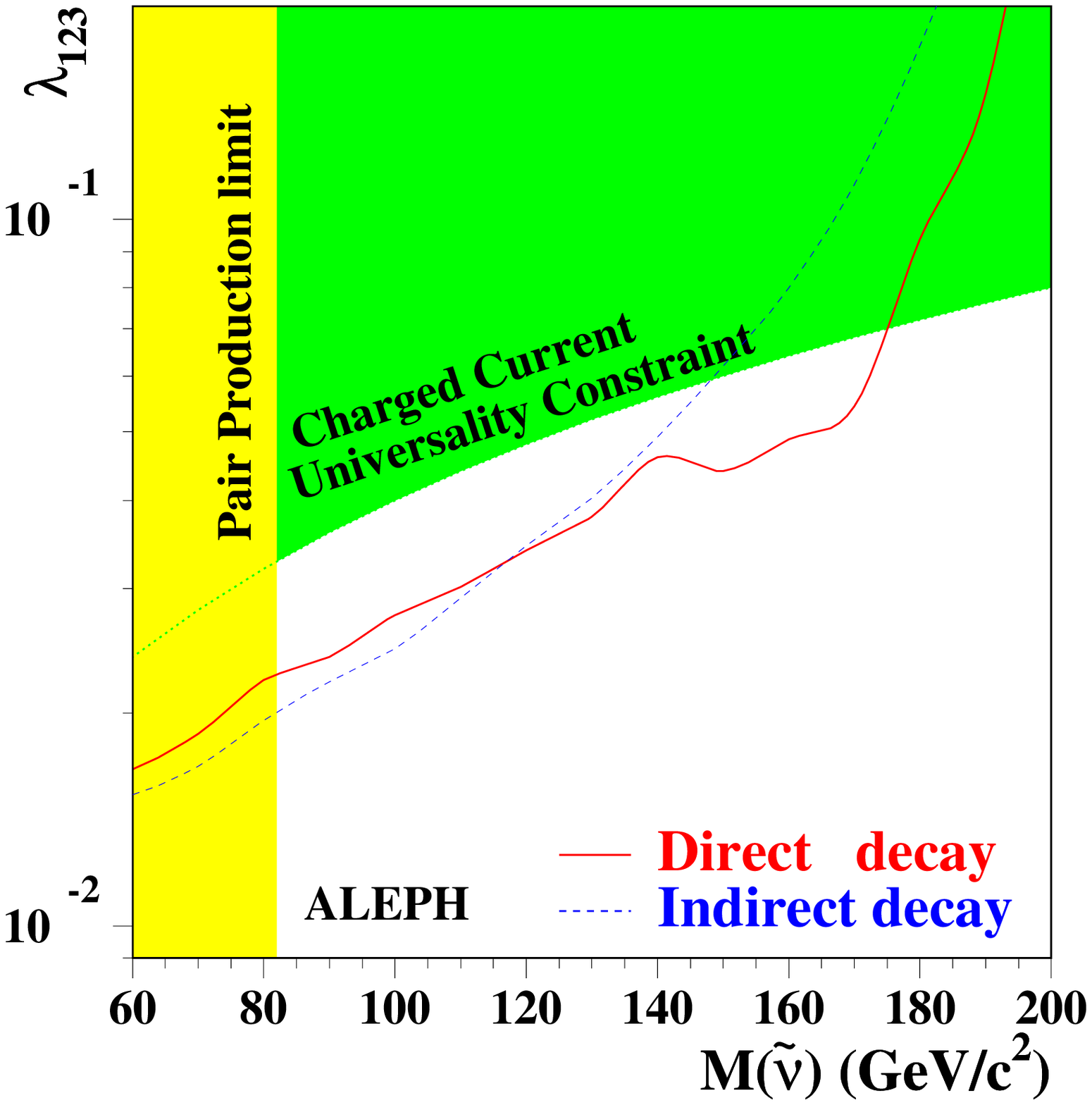,height=0.7\textheight}
\caption{\label{fig:123_excl}{   Observed 95\% CL upper limits on the 
$\lambda_{123}$ coupling for the direct and indirect decays.
The exclusions from the low energy bounds and the pair production
search are indicated.}} 
\end{center}
\end{figure}

\begin{figure}[!t]
\vspace{-0.7cm}
\begin{center}
\epsfig{file=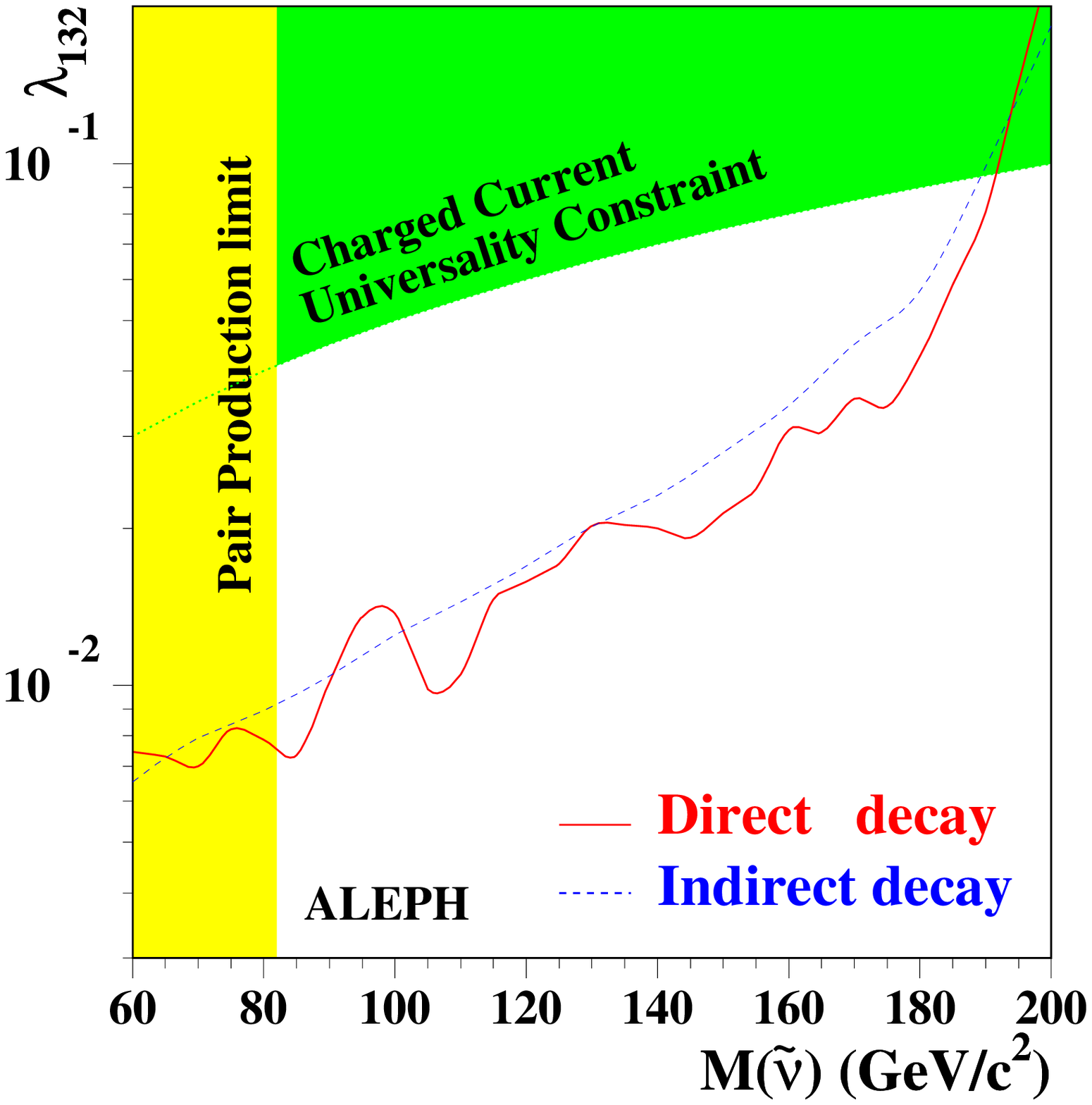,height=0.7\textheight}
\caption{\label{fig:132_excl}{  Observed 95\% CL upper limits on the 
$\lambda_{132}$ coupling for the direct and indirect decays.
The exculsions from the low energy bounds and the pair production
search are indicated.}} 
\end{center}
\end{figure}

\begin{figure}[!t]
\vspace{-1cm}
\begin{center}
\epsfig{file=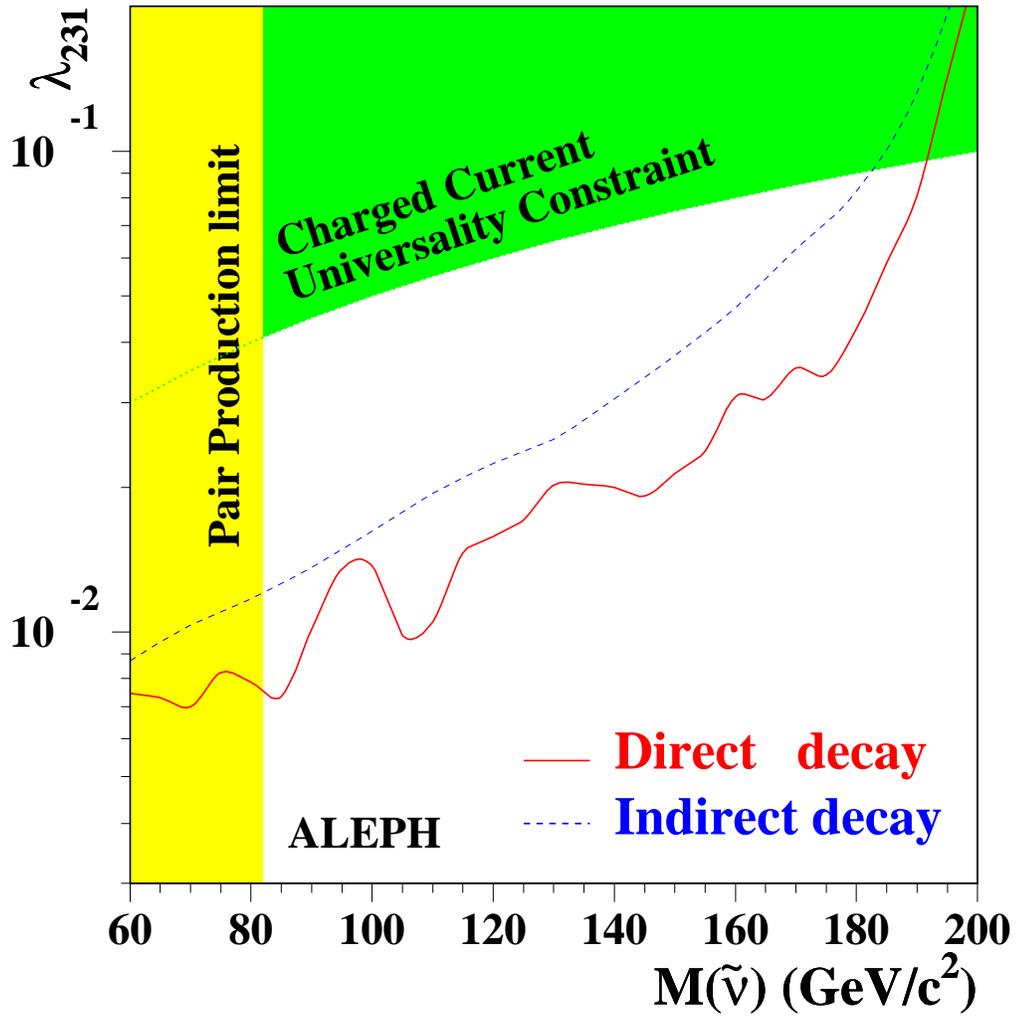,height=0.7\textheight}
\caption{\label{fig:231_excl}{  Observed 95\% CL upper limits on the 
$\lambda_{231}$ coupling for the direct ($\lambda_{1j2}$ analysis used)
and indirect decays.
The exclusions from the low energy bounds and the pair production
search are indicated.}} 
\end{center}
\end{figure}

\begin{figure}[!t]
\begin{center}
\epsfig{file=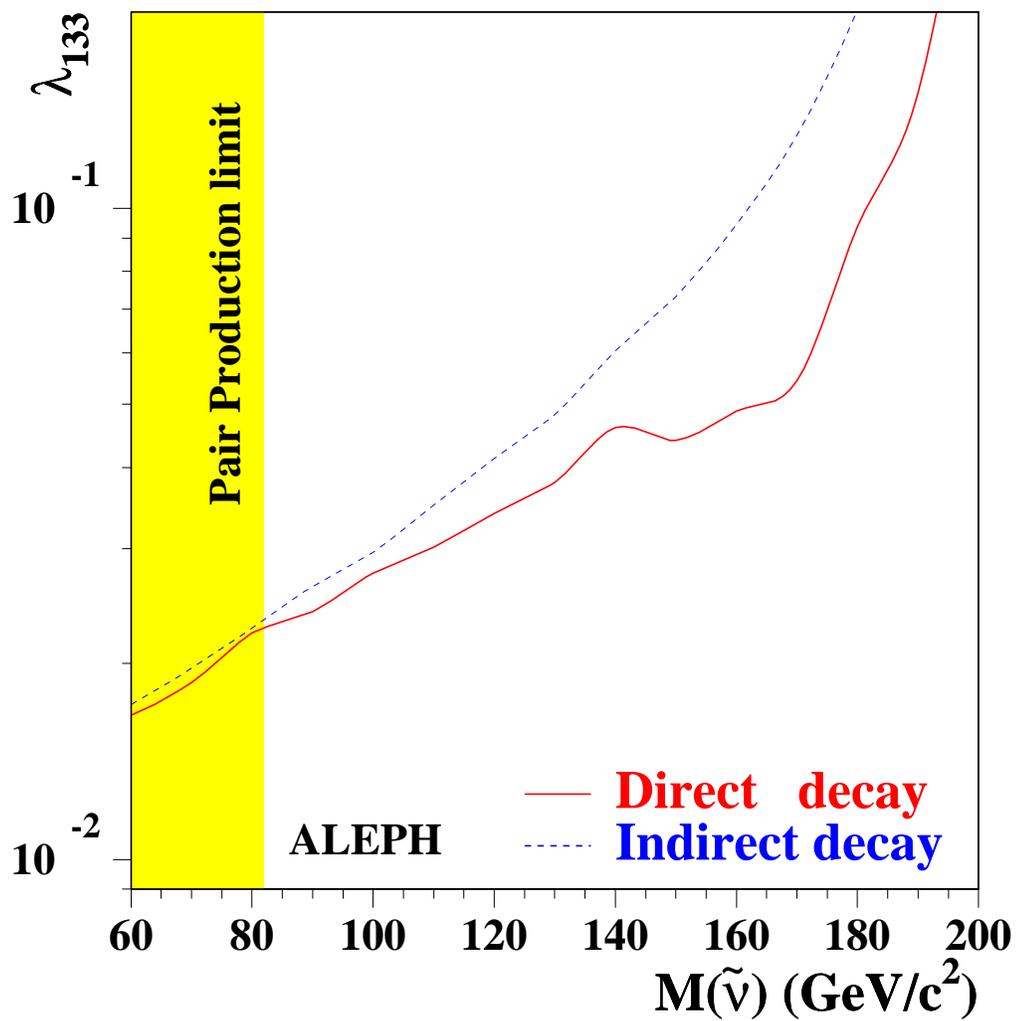,height=0.7\textheight}
\caption{\label{fig:133_excl}{  Observed 95\% CL upper limits on the 
$\lambda_{133}$ coupling for the direct and indirect decays.
The exclusion from the pair production search is indicated.}} 
\end{center}
\end{figure}

As the cross section for sneutrino production is a function of $\lambda$ 
and the sneutrino mass, the corresponding 95\% confidence level exclusions 
are presented in the $\lambda$ versus $M_{\mathrm{\tilde{\nu}}}$ plane in 
Figs.~\ref{fig:122_excl} to~\ref{fig:133_excl}. Also indicated on these 
figures are the exclusions obtained by the ALEPH search for pair production 
of sneutrinos decaying via a $LL\bar{E}$ coupling~\cite{rpv202} and the low 
energy bounds~\cite{leb}. 

Table~\ref{tab:lam_excl} compares the 95\% CL upper limits obtained on the 
couplings (direct and indirect decays), for an assumed sneutrino mass of 
$100\,\gevcc$, with the low energy bounds. Except for the $\lambda_{133}$ 
coupling, which already has a stringent upper limit from the 
$\nu_{\rm e}$ mass constraint, significantly improved upper limits are 
obtained on the other couplings for sneutrino masses up to $\sim 190\,\gevcc$.
 
\begin{table}[!t]
\caption{\ninerm The 95\% CL upper limits obtained on the various $\lambda$ 
couplings assuming a sneutrino mass of 100\,\gevcc. The last column shows 
the corresponding low energy bounds~\cite{leb}. In parentheses are indicated 
the sneutrino mass up to which these analyses improve upon the low energy 
bounds.
\newline }
\label{tab:lam_excl}
\centering
\begin{tabular}{|c||c|c|c|}\hline
$\lambda$ Coupling & \multicolumn{3}{|c|}{95\% CL Upper Limit}\\ 
                   & Direct Decay  & Indirect Decay    & Low Energy Bound\\    
\hline
  122              & 0.014~~~($189\,\gevcc$)    & 0.007~~~($190\,\gevcc$) & 0.04  \\ 
  123              & 0.028~~~($174\,\gevcc$)    & 0.025~~~($147\,\gevcc$) & 0.04  \\ 
  132              & 0.014~~~($191\,\gevcc$)    & 0.012~~~($187\,\gevcc$) & 0.05  \\ 
  133              & 0.028~~~(~~~~~~~-~~~~~~~)             & 0.029~~~(~~~~~~~-~~~~~~~)            & 0.003 \\ 
  231              & 0.014~~~($191\,\gevcc$)    & 0.016~~~($189\,\gevcc$) & 0.05  \\
\hline 
\end{tabular}
\end{table}

\section{Summary}
A number of searches sensitive to $R$-parity violating production and decay 
of single sneutrinos have been presented. These searches find no evidence 
for $R$-parity violated supersymmetry in the ALEPH data collected at 
$\sqrt{s}= 189$ to 209\,GeV. Within the framework of the MSSM, assuming that 
a single $\lambda$ coupling dominates and that the sneutrinos are degenerate
in mass, limits on the various $\lambda$ couplings as a function of the 
assumed sneutrino mass have been derived. These searches improve by up to 
a factor five upon the existing 95\% CL upper limits on four of the $\lambda$ 
couplings for sneutrino masses up to 190\,\gevcc, and are valid independent 
of whether the sneutrino decays directly or via the lightest neutralino. This 
paper is the first publication of such an analysis at LEP.

\section{Acknowledgements}
It is a pleasure to congratulate our colleagues from the accelerator divisions
for the successful operation of LEP at high energy. We would like to express 
our gratitude to the engineers and support people at our home institutes 
without whose dedicated help this work would not have been possible. 
Those of us from non-member states wish to thank CERN for its hospitality
and support.

\acrodef{LSP}{Lightest Supersymmetric Particle}
\acrodef{MSSM}{Minimal Supersymmetric Standard Model}
\acrodef{SM}{Standrad Model}
\acrodef{SUSY}{Supersymmetry}


\newpage

\begin{center}
{\Large \bf {Appendix}}
\end{center}
\vspace {-0cm}
\begin{table}[!h]
Table A1: Selection cuts for direct sneutrino decays via the
$\lambda_{122}$ and $\lambda_{132}$ couplings.
\newline\newline
\centering
\begin{tabular}{|c|c|} \hline 
\multicolumn{2}{|c|}{$E_{\mathrm{vis}}/\sqrt{s}>50\%$} \\
\multicolumn{2}{|c|}{$N_{\mathrm{e}}=1$} \\
\hline
{$N_{\mathrm{ch}}=2$}                      & {$N_{\mathrm{ch}}=3$}          \\
{$P_{\mathrm{T}}<8.5\,\gevc$}               & {$P_{\mathrm{T}}<10\,\gevc$}      \\
{$|\cos\theta_{\mathrm{e}}|<0.8$}          & {$|\cos\theta_{\mathrm{e}}|<0.9$} \\
{$N_{\mathrm{\mu}}=1$,~opp. charge to $e$} & {$N_{\mathrm{\mu}}=1$~or~2}  \\
{$|\cos\theta_{\mathrm{e\mu}}|<0.99$}      & {$|\cos\theta_{\mathrm{\mu t}}|<0.9$} \\
                                           & {$|M_{\mathrm{\mu t}}-M_{\mathrm{Z}}|>4\,\gevcc$} \\
\hline
\multicolumn{2}{|c|}{sliding mass cut: $|M_{\mathrm{e^\pm \mu^\mp}}-M_{\mathrm{\snu}}|<(M_{\mathrm{\snu}}/20-0.6)\,\gevcc$} \\
\hline
\end{tabular}
\end{table}
\vspace{2cm}

\vspace {-1cm}
\begin{table}[!h]
Table A2: Selection cuts for direct sneutrino decays via the
$\lambda_{123}$ and $\lambda_{133}$ couplings.
\newline\newline
\centering
\begin{tabular}{|c|}\hline
$N_{\mathrm{ch}}=$ 3~or~5,     \\
$0<N_{\mathrm{e}}<3, N_{\mathrm{\mu}}<2$           \\                       
$P_{\mathrm{T}}>6\,\gevc$               \\
$25\% <E_{\mathrm{vis}}/\sqrt s<90\%$     \\
$|\cos\theta_{\mathrm{miss}}|<0.98$       \\
\hline
$E_{\mathrm{emax}}>20~\mathrm{GeV}$ \\
$|\cos \theta_{\mathrm{emax}}|<0.9$ \\
\hline
2 $\tau$'s with opposite charge \\
1/1 or 1/3 prongs \\
$|\cos \theta_{\mathrm{\tau_1 \tau_2}}|<0.95$ \\
$P_{\mathrm{T(max E jet)}}>5\,\gevc$ \\
$|\cos \theta_{\mathrm{emax^\pm/\tau^\pm}}|>0.9$ \\
\hline
 sliding cuts:  \\
$f(M_{\mathrm{\snu}})<|\cos \theta_{\mathrm{emax^\pm/\tau^\mp}}|<f^\prime(M_{\mathrm{\snu}})$ \\
$f^{\prime\prime}(M_{\mathrm{\snu}})<E_{\mathrm{emax}}<f^{\prime\prime\prime}(M_{\mathrm{\snu}})$ \\
\hline
\end{tabular} \\
\end{table}
\vspace{0cm}

\begin{table}[!b]
Table A3: Selection cuts for indirect sneutrino decays via the
$\lambda_{122}$ coupling.
\newline\newline
\centering
\begin{tabular}{|c|c|c|} \hline 
\multicolumn{2}{|c|}{$N_{\mathrm{ch}}=3$}   \\
\hline
{$N_{\mathrm{\mu}}=3$}                         & {$N_{\mathrm{\mu}}=2$} \\
{$P_{\mathrm{T}}>4\,\gevc$}                & {$P_{\mathrm{T}}>6\,\gevc$} \\
{$E_{\mathrm{vis}}/\sqrt{s}<80\%$}      & {$E_{\mathrm{vis}}/\sqrt{s}<70\%$}  \\
{$|\cos\theta_{\mathrm{miss}}|<0.95$}  & {$|\cos\theta_{\mathrm{miss}}|<0.98$} \\
{$|M_{\mathrm{\mu\mu}}-M_{\mathrm{Z}}|>4\,\gevcc$}  & {$|\cos\theta_{\mathrm{t}}|<0.98$} \\

\hline
\end{tabular}
\end{table}

\begin{table}[!t]
Table A4: Selection cuts for indirect sneutrino decays via the
$\lambda_{132}$ coupling.
\newline\newline
\centering
\begin{tabular}{|c|}\hline
$N_{\mathrm{ch}}=3$ \\
$N_{\mathrm{e}}<2$, $N_{\mathrm{\mu}}>1$                \\                       
$P_{\mathrm{T}}>5\,\gevc$                   \\
$E_{\mathrm{vis}}/\sqrt s<80\%$     \\
$|\cos\theta_{\mathrm{miss}}|<0.95$      \\
acopJet$<170^\circ$ \\
\hline
\end{tabular}
\end{table}

\begin{table}[!t]
Table A5: Selection cuts for indirect sneutrino decays via the
$\lambda_{123}$ coupling.
\newline\newline
\centering
\begin{tabular}{|c|}\hline
$N_{\mathrm{ch}}=$ 3 or 5 \\
$N_{\mathrm{e}}+N_{\mathrm{\mu}}<4$                   \\                       
$P_{\mathrm{T}}>7\,\gevc$                     \\
$10\% <E_{\mathrm{vis}}/\sqrt s<60\%$      \\
$|\cos\theta_{\mathrm{miss}}|<0.9$         \\
acopJet$<170^\circ$             \\
$P_{\mathrm{T(max E jet)}}>5\,\gevcc$          \\
$y_{23}>2.5~10^{-4}$              \\
$|\cos\theta_{\mathrm{jet1,jet2}}|<0.95$    \\
\hline
\end{tabular}
\end{table}

\begin{table}[!t]
Table A6: Selection cuts for indirect sneutrino decays via the
$\lambda_{133}$ coupling. 
\newline\newline
\centering
\begin{tabular}{|c|c|}\hline

\multicolumn{2}{|c|} {$N_{\mathrm{ch}}=3$~or~5}  \\
\multicolumn{2}{|c|} {$N_{\mathrm{e}} + N_{\mathrm{\mu}}<3$}                           \\                       
\multicolumn{2}{|c|} {$|\cos\theta_{\mathrm{miss}}|<0.95$}                \\
\multicolumn{2}{|c|} {acopJet$<160^\circ$}    \\
\hline
$N_{\mathrm{ch}}=3$                     &  $N_{\mathrm{ch}}=5$   \\
$P_{\mathrm{T}}>10\,\gevc$               &  $P_{\mathrm{T}}>7\,\gevc$         \\
$E_{\mathrm{vis}}/\sqrt{s} <50\%$       &  $E_{\mathrm{vis}}/\sqrt{s} <60\%$ \\  
$|\cos \theta_{\mathrm{jet-jet}}| <0.9$ &                                 \\
\hline
\end{tabular}
\end{table}

\begin{table}[!t]
Table A7: Selection cuts for indirect sneutrino decays via the
$\lambda_{231}$ coupling.
\newline\newline
\centering
\begin{tabular}{|c|}\hline
$N_{\mathrm{ch}}=3$ \\
$N_{\mathrm{e}}>0$, $N_{\mathrm{\mu}}>0$               \\                       
$P_{\mathrm{T}}>7\,\gevc$           \\
$E_{\mathrm{vis}}/\sqrt s<75\%$      \\
$|\cos\theta_{\mathrm{miss}}|<0.94$       \\
acopJet$<170^\circ$ \\
\hline
\end{tabular}
\end{table}

\newpage

\end{document}